\documentclass[aps,prd,reprint,preprintnumbers,superscriptaddress,nofootinbib,floatfix]{revtex4-1}

\usepackage{amsmath,amssymb,amsfonts,bm}
\usepackage{graphicx}
\usepackage{xcolor}
\usepackage{slashed}
\usepackage{placeins}
\usepackage{tabularx}
\usepackage[colorlinks=true,linkcolor=blue,citecolor=blue,urlcolor=blue,filecolor=blue]{hyperref}
\usepackage{comment}
\usepackage{subfigure}
\usepackage[normalem]{ulem}

\newcommand{\UNISA}{\affiliation{Dipartimento di Fisica ``E.R.\ Caianiello'', Universit\`a degli Studi di Salerno,\\ Via Giovanni Paolo II, 132 - 84084 Fisciano (SA), Italy}}
\newcommand{\INFN}{\affiliation{Istituto Nazionale di Fisica Nucleare - Gruppo Collegato di Salerno - Sezione di Napoli,\\ Via Giovanni Paolo II, 132 - 84084 Fisciano (SA), Italy}}

\begin{document}

\title{Microscopic primordial black holes as macroscopic dark matter from large extra dimensions}

\preprint{IPPP/26/33}

\author{Giuseppe Filiberto Vitale}
\email{givitale@unisa.it}
\UNISA \INFN

\author{Gaetano Lambiase}
\email{glambiase@unisa.it}
\UNISA \INFN

\author{Tanmay Kumar Poddar}
\email{tanmay.k.poddar@durham.ac.uk}
\affiliation{Institute for Particle Physics Phenomenology, Department of Physics, Durham University,\\
Durham, DH1 3LE United Kingdom}

\author{Luca Visinelli}
\email{lvisinelli@unisa.it}
\UNISA \INFN

\begin{abstract}
We study the coupled cosmological evolution of primordial black holes (PBHs) and radiation in the Arkani-Hamed-Dimopoulos-Dvali (ADD) framework with $n$ large extra dimensions and a fundamental gravity scale $M_\star$ at the TeV scale. For PBHs with horizon radius smaller than the compactification scale,  the higher-dimensional geometry implies a larger horizon size at fixed mass and therefore a suppressed Hawking temperature. As a result, radiation accretion can overcome evaporation in the early Universe and drive a ``runaway'' phase of rapid mass growth. By numerically solving the coupled mass and energy-density evolution equations, we show that for $n \geq 2$ initially microscopic PBHs with initial mass $M_i \gtrsim 10^{12}\,$g can grow by many orders of magnitude and potentially reach macroscopic, even solar-mass, scales by matter-radiation equality. We determine the critical initial abundance $\beta_{\rm crit}$ required for PBHs to account for the observed dark matter density and find that extra dimensions dramatically lower this threshold, allowing viable scenarios with $\beta_{\rm crit}\sim 10^{-44}$. This identifies a previously unexplored region of parameter space in which the dark matter abundance is achieved through dynamical mass growth rather than large initial collapse fractions.
\end{abstract}

\maketitle

\section{Introduction}

Understanding the nature of dark matter (DM) and the origin of the hierarchy problem between the electroweak and Planck scales remain two of the most pressing challenges in fundamental physics. While these questions are often addressed independently, frameworks that modify gravity at short distances offer the intriguing possibility of linking them within a unified picture. Among the most compelling proposals is the Arkani-Hamed-Dimopoulos-Dvali (ADD) scenario~\cite{Arkani-Hamed:1998jmv, Antoniadis:1998ig, Randall:1999ee, Randall:1999vf}, in which the fundamental scale of gravity, $M_\star$, is lowered to the TeV range due to the presence of $n$ large compact extra dimensions. In this framework, the observed weakness of four-dimensional gravity emerges as a macroscopic consequence of gravitational flux dilution into the higher-dimensional bulk, while Standard Model (SM) fields remain confined to a three-dimensional brane. 

Existing constraints on large extra dimensions (LEDs) arise from a wide range of laboratory, astrophysical, and cosmological probes~\cite{ParticleDataGroup:2024cfk}. In such scenarios, modifications to Newtonian gravity can be described approximately by a Yukawa-type potential generated by higher dimensional gravitational modes, and are therefore constrained by short-distance tests including E\"ot-Wash~\cite{Adelberger:2002ic,Antoniadis:1998ig} and atomic~\cite{Dzuba:2022xvo} experiments. Complementary bounds follow from collider searches for missing energy events and deviations in high energy scattering~\cite{Franceschini:2011wr}, from neutrino oscillation data in models where sterile neutrinos behave as new massive states~\cite{MINOS:2016vvv,deGiorgi:2025xgp}, and from neutron star and supernova production of massive modes whose decays may yield observable signals~\cite{Hannestad:2003yd,Fermi-LAT:2012zxd,Barger:1999jf}. Further cosmological limits arise because excessive production of these modes in the early Universe could lead to an early matter-dominated era and hence an unacceptably small age of the Universe~\cite{Fairbairn:2001ct}.

Black holes (BHs) provide a powerful probe of such modifications of gravity, as their properties directly reflect the underlying gravitational dynamics and spacetime dimensionality. In particular, BHs with horizon radii smaller than the compactification scale $R$ probe the $(4+n)$-dimensional regime, where their mass-radius relation, temperature, and evaporation rates differ qualitatively from the familiar four-dimensional Schwarzschild behavior~\cite{Myers:1986un, Argyres:1998qn, Giddings:2001bu, Dimopoulos:2001hw, Emparan:2008eg, Sirletti:2025xfv}. These deviations become especially significant for microscopic BHs formed in the early Universe. This perspective has recently been explored more broadly, with compact objects probing dark-sector physics and modified gravity~\cite{Brax:2026cmh}.

Primordial black holes (PBHs)~\cite{Zeldovich:1967lct, Hawking:1971ei, Carr:1974nx, Carr:1975qj} have long been recognized as a viable and well-motivated DM candidate (see e.g.~\cite{Carr:2020xqk, Green:2020jor, Carr:2020gox, Villanueva-Domingo:2021spv, Escriva:2022duf} for reviews). PBHs can form from the collapse of large density fluctuations in the early Universe, with particularly efficient production possible during periods of reduced pressure support, such as the QCD phase transition, where the equation of state temporarily softens~\cite{Jedamzik:1996mr, Byrnes:2018clq}. PBHs have also been proposed as seeds for early structure formation and for explaining the massive high-redshift galaxies observed by the James Webb Space Telescope, particularly in scenarios where rapid mass growth is possible~\cite{Yuan:2023bvh, Cai:2023ptf, Li:2023zva}, potentially alleviating tensions with early structure formation~\cite{Iocco:2024rez}. Unlike particle DM, PBHs probe directly the physics of the primordial Universe through their formation and subsequent evolution, and do not require extensions of the SM. In standard four-dimensional cosmology, however, the evolution of PBHs is strongly constrained. Objects lighter than $\sim 10^{15}\,$g evaporate efficiently via Hawking radiation, while heavier PBHs are subject to a wide range of observational bounds across different mass scales~\cite{Carr:2009jm, Carr:2020erq, Carr:2020mqm, Coogan:2020tuf, Serpico:2020ehh, Racco:2022bwj, Keith:2022sow}. As a result, only relatively narrow mass windows remain viable for PBH to constitute the entirety of DM, although extended mass functions and non-standard cosmological histories can relax these constraints~\cite{Franciolini:2021xbq, Mazde:2022sdx, Boccia:2024nly, Hooper:2020evu}.\footnote{See however the light mass window accessible via memory burden effects~\cite{Dvali:2018xpy, Dvali:2020wft, Dvali:2024hsb, Zantedeschi:2024ram, Chianese:2024rsn, Leontaris:2025piz, Ettengruber:2026jrc, Leontaris:2026kvu, Ahmed:2026pjd}.} Recent developments have further emphasized both the role of early-Universe dynamics in shaping PBH abundances and the sensitivity of their phenomenology to beyond SM particle content and modified evaporation channels~\cite{Calza:2023rjt, Auffinger:2020afu, Yin:2026hvw}. In parallel, gravitational wave observations have emerged as a powerful and complementary probe of PBHs through their merger rates and mass distributions~\cite{Nakamura:1997sm, Bird:2016dcv, Clesse:2016vqa, Sasaki:2016jop, Ali-Haimoud:2017rtz, Raidal:2018bbj, Belotsky:2018wph, Mukherjee:2021ags, Boybeyi:2024aax, Yuan:2024yyo, Domenech:2026nun, Haque:2026yum}.

A key limitation of the standard picture is that accretion of radiation in the early Universe is typically inefficient~\cite{Ali-Haimoud:2016mbv, DeLuca:2020fpg}. The combination of relativistic pressure and rapid cosmic expansion prevents the formation of a sustained inflow, so that PBHs evolve approximately as isolated objects with nearly constant mass until evaporation becomes important~\cite{Carr:1974nx, Niemeyer:1999ak, Bean:2002kx, Ricotti:2007au}. Consequently, significant mass growth from microscopic initial seeds is not expected in conventional cosmology. This situation can change qualitatively in the presence of extra dimensions. In the ADD framework, BHs with horizon radii below the compactification scale have larger radii at fixed mass, implying lower Hawking temperatures and suppressed evaporation rates compared to the four-dimensional case~\cite{Argyres:1998qn, Anchordoqui:2024dxu, Anchordoqui:2025xug}. At the same time, the enlarged horizon enhances the effective capture cross-section for ambient radiation. As a result, accretion from the primordial plasma can compete with, and in some regimes dominate over, Hawking evaporation~\cite{Carr:1974nx, MacGibbon:1991tj, Bean:2002kx, Ricotti:2007au, Friedlander:2022ttk}. When this condition is satisfied, PBHs enter a regime of ``runaway'' accretion, defined by $\dot M_{\rm acc} \gg |\dot M_{\rm evap}|$ and a growth timescale shorter than the Hubble time. In this phase, PBHs efficiently extract energy from the surrounding radiation bath and grow rapidly in mass, significantly modifying both their evolution and the global energy budget of the Universe. This runaway behavior is a distinctive feature of the higher-dimensional scenario and does not arise in standard four-dimensional cosmology under comparable conditions.

In this work, we investigate the possibility of runaway accretion in detail by studying the coupled cosmological evolution of PBHs and the radiation background within the ADD framework. Starting from initial conditions at formation, we track the mass growth and energy density of PBHs across cosmic time, with particular attention to the transition between higher-dimensional and effectively four-dimensional regimes~\cite{Argyres:1998qn, Emparan:2008eg}. We show that, for $n \geq 2$, the suppression of Hawking evaporation combined with enhanced accretion can trigger a prolonged phase of growth, allowing initially microscopic PBHs to reach macroscopic masses, potentially as large as $\mathcal{O}(M_\odot)$, by matter-radiation equality. This demonstrates that extra-dimensional effects can provide a viable mechanism through which microscopic seeds evolve into cosmologically relevant objects. In particular, this scenario provides a natural formation channel for solar mass PBHs, including in the sub-solar mass range where standard stellar evolution cannot produce BH remnants~\cite{LIGOScientific:2021job, Magaraggia:2026jhk, Haque:2026yum}.

A key outcome of this study is the determination of the critical initial abundance, $\beta_{\rm crit}$, required for PBHs to account for the observed DM density~\cite{Carr:1975qj}. We find that extra dimensions dramatically lower this threshold, with viable scenarios extending to $\beta \sim 10^{-44}$, far below standard expectations. This identifies a previously unexplored region of parameter space in which the DM abundance is achieved through dynamical mass growth rather than large initial collapse fractions. The present work establishes the cosmological viability of this mechanism. A distinctive feature of this scenario is that late time observational constraints must be reinterpreted through the non-trivial mapping between the formation mass and the present day PBH mass induced by early-Universe accretion.

The paper is organized as follows. In Sec.~\ref{sec:methods}, we review the properties of BHs in the ADD framework and establish the formalism used to describe their evolution across higher-dimensional and four-dimensional regimes. In Sec.~\ref{sec:evolution}, we present the coupled evolution equations and numerical analysis for PBHs in large extra-dimensional framework. In Sec.~\ref{sec:results}, we discuss the resulting DM abundance and determine the critical initial fraction. In Sec.~\ref{sec:discussion}, we interpret our results in light of current observational constraints. Finally, Sec.~\ref{sec:conclusions} summarizes our findings and outlines future directions. We adopt the convention $\hbar = c = k_B = 1$ throughout the paper, unless stated otherwise.

\section{Methods}
\label{sec:methods}

\subsection{Black holes in large extra dimensions}
\label{sec:ADD_BH}

In models with LEDs, the fundamental $(4+n)$-dimensional Planck scale $M_\star$ is related to the observed four-dimensional Planck mass $m_{\rm Pl}$ through the compactification radius $R$ of the extra dimensions, given by~\cite{Myers:1986un,Argyres:1998qn}
\begin{equation}
    \label{eq:R_ADD}
    R = \frac{1}{2\pi}\left(\frac{m_{\rm Pl}^2}{M_\star^{\,n+2}}\right)^{1/n}\,,
\end{equation}
where $n$ denotes the number of extra spatial dimensions. At distances $r < R$, gravity propagates in the full $(4+n)$-dimensional spacetime and the gravitational potential scales as $V(r) \propto r^{-(1+n)}$, while for $r > R$ the standard four-dimensional behavior is recovered. A BH with horizon radius smaller than the compactification scale probes the higher-dimensional gravitational regime. A non-rotating BH of mass $M_{\rm BH}$ formed in this regime ($r_h < R$) has a horizon radius~\cite{Argyres:1998qn}
\begin{equation}
    \label{eq:rh_LED}
    r_h^{\rm LED} = \frac{a_n}{M_\star}\left(\frac{M_{\rm BH}}{M_\star}\right)^{\frac{1}{n+1}}\,,
\end{equation}
where the dimensionless coefficient
\begin{equation}
    \label{eq:an}
    a_n = \left[\frac{8\Gamma((n+3)/2)}{(n+2)\pi^{(n+1)/2}}\right]^{\frac{1}{n+1}}\,,
\end{equation}
arises from the $(4+n)$-dimensional Schwarzschild solution. For sufficiently large masses, the horizon radius exceeds the compactification scale, $r_h > R$, and the geometry becomes effectively four-dimensional. In this limit, the standard Schwarzschild radius applies~\cite{Hawking:1974rv},
\begin{equation}
    \label{eq:rh_GR}
    r_h^{\rm 4D} = \frac{2M_{\rm BH}}{m_{\rm Pl}^2}\,.
\end{equation}

To describe BHs evolving across both regimes, we adopt a piecewise prescription for the horizon radius,
\begin{equation}
    \label{eq:rh_piecewise}
    r_h(M_{\rm BH})=
    \begin{cases}
        \frac{a_n}{M_\star}\left(\frac{M_{\rm BH}}{M_\star}\right)^{\frac{1}{n+1}}\,,
        & M_{\rm BH} \leq M_{\rm 4D}\,,\\[10pt]
        \dfrac{2M_{\rm BH}}{m_{\rm Pl}^2}\,,
        & M_{\rm BH} > M_{\rm 4D}\ \text{or}\ n=0\,,
    \end{cases}
\end{equation}
where the transition mass $M_{\rm 4D}$ is determined by the condition that the four-dimensional Schwarzschild radius equals the compactification scale, $r_h^{\rm 4D} \sim R$.\footnote{In reality, the transition between the higher-dimensional and four-dimensional regimes is expected to be smooth. The piecewise prescription adopted here provides a controlled interpolation that reproduces the correct asymptotic scaling in both limits, although a fully consistent treatment would require a smooth higher-dimensional solution across the compactification scale.} This yields
\begin{equation}
    \label{eq:M4D}
    	M_{\rm 4D} = \dfrac{1}{4\pi}\dfrac{m_{\rm Pl}^2}{M_\star}\,\left(\dfrac{m_{\rm Pl}}{M_\star}\right)^{2/n}\,.
\end{equation}
This mass scale marks the transition between higher-dimensional and effectively four-dimensional. Representative values of $M_{\rm 4D}$ for different choices of the fundamental $(4+n)$-dimensional gravity scale $M_\star$ are reported in Table~\ref{tab:critical_masses}.

\begin{table}[ht]
    \centering
    \renewcommand{\arraystretch}{1.4}
    \begin{tabular}{c @{\hspace{2em}} c @{\hspace{2em}} c @{\hspace{2em}} c}
\hline
        & \multicolumn{3}{c}{$M_{\rm 4D}$} \\
\hline
        $n$ & $M_\star = 10\,$TeV & $M_\star = 10^3\,$TeV & $M_\star = 10^5\,$TeV \\
\hline
        2 & $2.6 \times 10^{24}$\,g & $2.6 \times 10^{20}$\,g & $2.6 \times 10^{16}$\,g \\
        3 & $2.4 \times 10^{19}$\,g & $1.1 \times 10^{16}$\,g & $5.2 \times 10^{12}$\,g \\
        4 & $7.4 \times 10^{16}$\,g & $7.4 \times 10^{13}$\,g & $7.4 \times 10^{10}$\,g \\
        5 & $2.3 \times 10^{15}$\,g & $3.6 \times 10^{12}$\,g & $5.8 \times 10^9\phantom{0}$\,g \\
        6 & $2.3 \times 10^{14}$\,g & $4.9 \times 10^{11}$\,g & $1.0 \times 10^9\phantom{0}$\,g \\
\hline
    \end{tabular}
    \renewcommand{\arraystretch}{1}
    \caption{Transition mass $M_{\rm 4D}$ separating higher-dimensional and effectively four-dimensional BHs, for representative values of the fundamental scale $M_\star = 10, 10^3, 10^5$\,TeV, and different numbers of extra dimensions $n$.}
    \label{tab:critical_masses}
\end{table}

The thermodynamic evolution of BHs is governed by the Hawking temperature $T_H$. In the four-dimensional regime the temperature takes the familiar form
\begin{equation}
    T_H^{\rm 4D} = \frac{1}{4\pi\, r_h^{\rm 4D}}\,,
\end{equation}
whereas in the higher-dimensional regime ($r_h < R$) it becomes~\cite{Argyres:1998qn}
\begin{equation}
    \label{eq:HawkingLED}
    T_H^{\rm LED} = \frac{n+1}{4\pi\, r_h^{\rm LED}}\,.
\end{equation}
Because the horizon radius grows more rapidly with mass in higher dimensions, the corresponding Hawking temperature is reduced at fixed mass with respect to the standard four-dimensional case. This modified temperature-mass relation also leads to a different scaling of the evaporation timescale. In the higher-dimensional regime, the mass loss rate due to Hawking radiation can be estimated as
\begin{equation}
    \label{eq:dMdt_evap0}
    \frac{{\rm d}M_{\rm BH}}{{\rm d}t}\bigg|_{\rm evap} = - \epsilon_n \, A_{(4+n)} \, T_H^{\,4+n}\,,
\end{equation}
where $\epsilon_n$ encodes graybody factors and the effective number of radiated degrees of freedom~\cite{Harris:2003eg, Kanti:2004nr, Barman:2019vst}. The horizon area is $A_{(4+n)} = \Omega_{(2+n)} r_h^{2+n}$, with $\Omega_{(2+n)} = 2\pi^{(3+n)/2}/\Gamma((3+n)/2)$. The scalings $A_{(4+n)} \propto r_h^{2+n}$ and $T_H \propto r_h^{-1}$ imply $A_{(4+n)} T_H^{4+n} \propto T_H^2$, independently of the number of extra dimensions $n$. It is therefore convenient to rewrite Eq.~\eqref{eq:dMdt_evap0} as
\begin{equation}
    \label{eq:dMdt_evap}
    \frac{{\rm d}M_{\rm BH}}{{\rm d}t}\bigg|_{\rm evap} \equiv -\alpha(n)\,T_H^2\,,
\end{equation}
where the coefficient $\alpha(n)$ absorbs the numerical prefactors, graybody factors, and the effective number of degrees of freedom. We adopt the values for $\alpha(n)$ from Table~II of Ref.~\cite{Friedlander:2022ttk}, computed in the approximation of massless emitted particles and including both brane and bulk emission channels.

Substituting the higher-dimensional Hawking temperature in Eq.~\eqref{eq:dMdt_evap}, we obtain
\begin{equation}
    \frac{{\rm d}M_{\rm BH}}{{\rm d}t}\bigg|_{\rm evap} = - \kappa_n \, M_\star^{2}\left(\frac{M_{\rm BH}}{M_\star}\right)^{-\frac{2}{n+1}}\,,
\end{equation}
where
\begin{equation}
    \kappa_n = \alpha(n) \,\left(\frac{n+1}{4\pi}\right)^{4+n}a_n^{-2}\,.
\end{equation}
Integrating the mass loss rate, we obtain the evaporation timescale
\begin{equation}
	\label{eq:evaporation}
	\tau_{\rm BH}(M_{\rm BH}) \propto \frac{n+1}{(n+3)\,\kappa_n}\;\frac{1}{M_\star}\left(\frac{M_{\rm BH}}{M_\star}\right)^{\frac{n+3}{n+1}}\,.
\end{equation}
This result generalizes the standard four-dimensional scaling $\tau_{\rm BH} \propto M_{\rm BH}^3$. In particular, the exponent $(n+3)/(n+1)$ is smaller than $3$ for $n \geq 1$, leading to a modified mass dependence of the lifetime. The overall lifetime can be either enhanced or reduced depending on the mass scale and the value of $M_\star$. The prefactor in Eq.~\eqref{eq:evaporation} is fixed by requiring consistency with the standard four-dimensional result, namely that a BH with mass $M_0 \simeq 5\times 10^{14}\,$g evaporates at the present epoch, $\tau_{\rm BH} \simeq \tau_U$. With this calibration, the higher-dimensional expression provides a consistent extension of the evaporation timescale across the two regimes.

\begin{figure}[hbt]
    \centering
    \includegraphics[width=\linewidth]{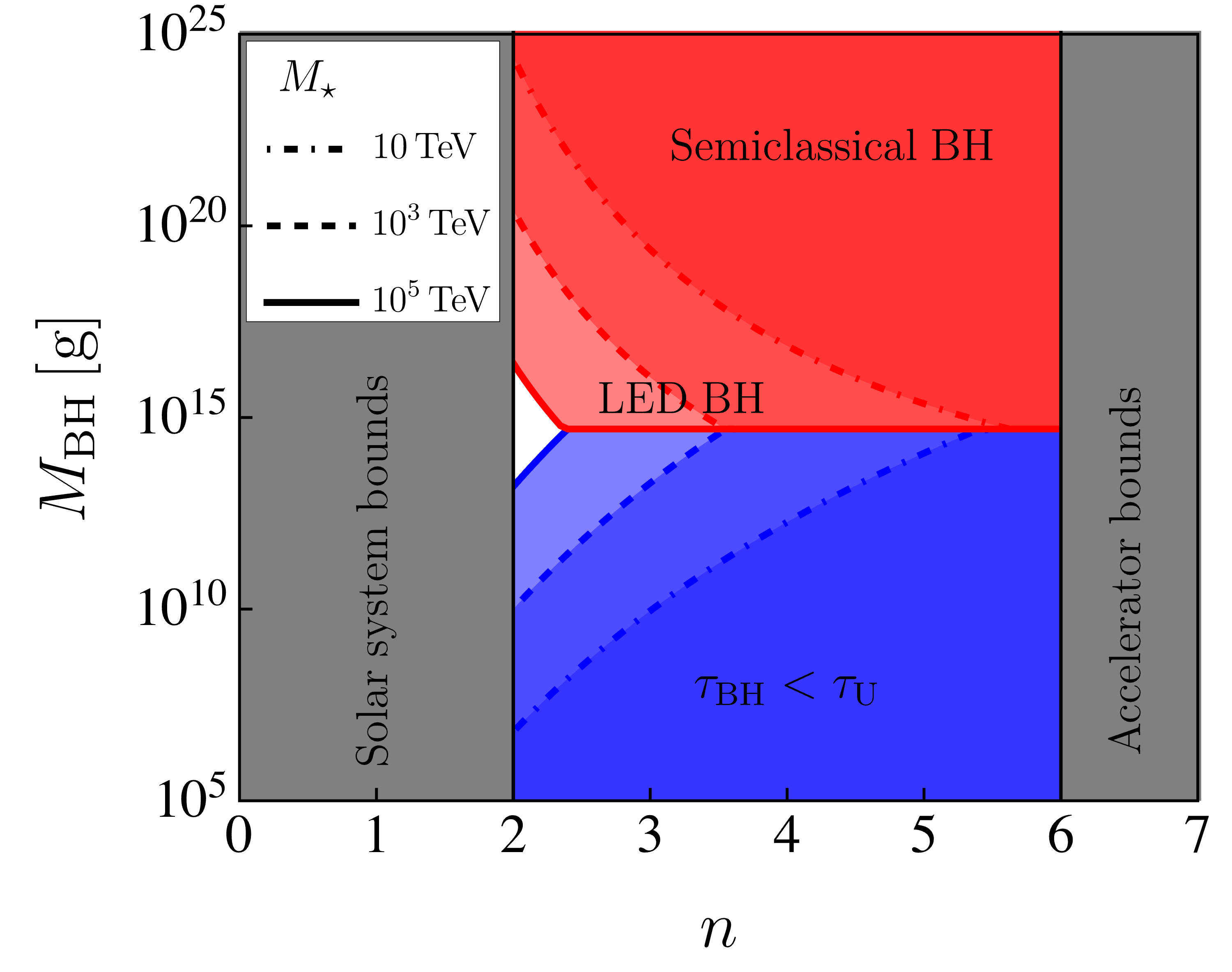}
    \caption{Phase diagram of BHs in the presence of $n$ LED, shown in the $(M_{\rm BH}, n)$ plane. The solid, dashed and dot-dashed lines marks the transition between the higher-dimensional regime ($r_h < R$) and the effective four-dimensional regime ($r_h > R$), corresponding to $M_{\rm BH} \simeq M_{\rm 4D}$. The red region indicates BHs in the effective four-dimensional (``semiclassical'') regime, $M_{\rm BH} > M_{\rm 4D}$, while the blue region corresponds to BHs with lifetimes shorter than the age of the Universe ($\tau_{\rm BH} < \tau_U$). Grey bands denote constraints from laboratory and astrophysical tests of gravity~\cite{Adelberger:2003zx, Arkani-Hamed:1998jmv}. The solid, dashed and dot-dashed curves show constant lifetime contours.}    
    \label{fig:BHplot}
\end{figure}

\begin{figure}[htb]
    \centering
    \includegraphics[width=\linewidth]{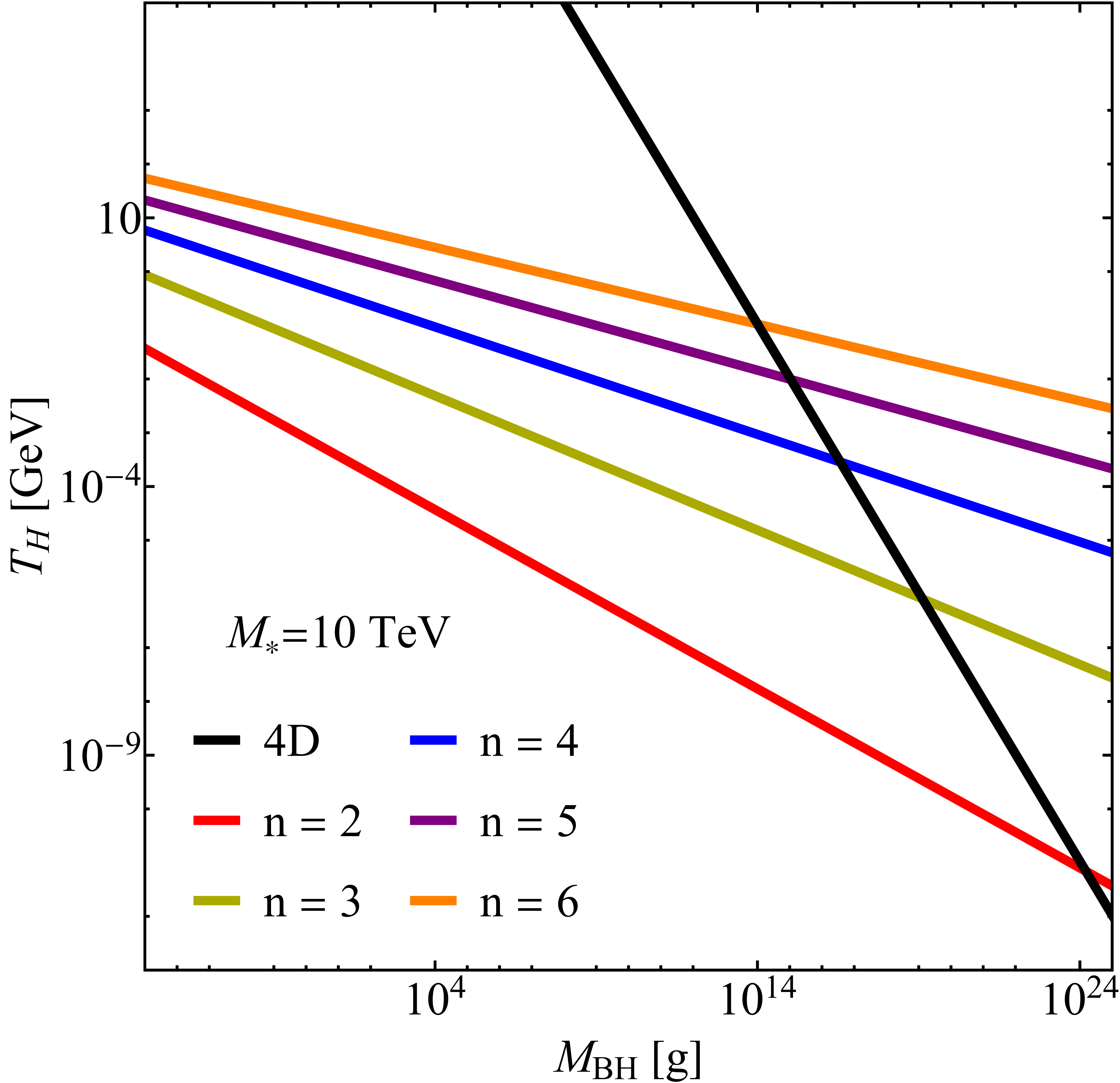}
    \caption{Hawking temperature of BHs as a function of mass for different numbers of extra dimensions $n$, assuming $M_\star = 10$\,TeV. Models with extra dimensions predict substantially lower Hawking temperatures compared to the standard four-dimensional case.}
    \label{fig:HawkingTemperature}
\end{figure}

The phase diagram of BHs in the presence of $n$ LED is illustrated in Fig.~\ref{fig:BHplot}, which summarizes the different regimes of BH evolution in the $(M_{\rm BH}, n)$ plane. The red shaded region labeled ``Semiclassical BH'' corresponds to $M_{\rm BH} > M_{\rm 4D}$ for different values of $M_\star$ in Eq.~\eqref{eq:M4D}. For sufficiently large values of $M_\star$, e.g. $M_\star = 10^5$\,TeV, the cosmologically relevant mass range lies entirely in the four-dimensional regime, so that the effects of extra dimensions become negligible. Further constraints on the model are obtained by equating the evaporation timescale in Eq.~\eqref{eq:evaporation} to the age of the Universe, $\tau_{\rm BH} \sim \tau_U$, yielding  the mass of BHs evaporating at the present epoch as
\begin{equation}
    \label{eq:Mevap_LED}
    M_{\rm evap} = M_0 \left(\frac{\tau_U}{\tau_{\rm BH}(M_0)}\right)^{\frac{n+1}{n+3}} \,,
\end{equation}
which defines the boundary between evaporating and long-lived BHs in the higher-dimensional regime shown in Fig.~\ref{fig:BHplot} (blue shaded region). The result for $n = 1$ labeled ``Solar system bounds'' is excluded by the fact that, in the original ADD model, if there were only one large extra dimension, its radius $R$ would be so large that ordinary Newtonian gravity would be altered at solar‑system scales~\cite{Arkani-Hamed:1998jmv}. Moreover, collider searches at the LHC place strong lower limits on the fundamental gravity scale in ADD models, effectively restricting large numbers of extra dimensions for TeV‑scale scenarios~\cite{Franceschini:2011wr}. The resulting temperature-mass relation is shown in Fig.~\ref{fig:HawkingTemperature}. In the higher-dimensional regime, the weaker dependence of the Hawking temperature on mass leads to significantly lower temperatures at fixed $M_{\rm BH}$ compared to the four-dimensional case. This suppression delays evaporation and enhances the relative importance of accretion, opening the possibility of a growth phase that is absent in standard four-dimensional cosmology under conventional assumptions~\cite{Friedlander:2022ttk}, although recent work suggests that general relativistic effects may enhance accretion even in the 4D case~\cite{Haque:2026vvp, Kallifatides:2026sik,Das:2025vts, Kalita:2025fcs}.

\subsection{Initial conditions and PBH formation}
\label{sec:initial_conditions}

After formation in the early Universe, the subsequent evolution of the PBH population is highly sensitive to the conditions at formation,
%since the accretion efficiency scales as $T^4/T_H^2$. 
which determine both the temperature of the radiation bath and the onset of the accretion phase. We assume that PBHs form through the gravitational collapse of sufficiently large primordial density fluctuations upon horizon re-entry during radiation domination. In this scenario, the formation time $t_i$ is related to the initial PBH mass $M_i$ through
\begin{equation}
    \label{eq:formation}
    t_i = \frac{M_i}{\gamma m_{\rm Pl}^2} \,,
\end{equation}
where $\gamma \sim 0.2$ parametrizes the collapse efficiency~\cite{Carr:1975qj}. This relation follows from the standard four-dimensional horizon-mass relation.
%Since the background cosmological expansion remains effectively four-dimensional in the ADD framework, this relation continues to hold in the presence of extra dimensions.
For some regions of the parameter space, however, the Hubble radius at formation may become smaller than the compactification scale, $H(t_i)^{-1}<R$. In this regime, the causal region relevant for PBH formation could probe gravity in higher dimensions, and the relation between the horizon mass and the formation time becomes model dependent. Modified PBH formation laws have indeed been derived in alternative extra-dimensional cosmologies, such as the Randall--Sundrum braneworld scenario~\cite{Guedens:2002km,Guedens:2002sd,Clancy:2003zd}. Throughout this work, we adopt Eq.~\eqref{eq:formation} as a phenomenological benchmark, corresponding to an effectively four-dimensional cosmological background with stabilized extra dimensions and SM fields confined to the brane~\cite{Hall:1999mk,Hannestad:2001jv}.

%The physical picture is that gravitational collapse can only occur once the Hubble horizon encloses a region whose density exceeds a critical threshold.
Prior to horizon entry, pressure gradients efficiently counteract gravitational contraction, preventing the collapse of overdense regions. During radiation domination, the Hubble rate is $H = 1/(2t)$ and the radiation energy density satisfies
\begin{equation}
    \label{eq:rhor}
    \rho_r = \frac{\pi^2}{30} g_\star(T) T^4\,,
\end{equation}
where $g_\star(T)$ accounts for the effective relativistic degrees of freedom at $T$.
%For the relativistic degrees of freedom at energies above the SM scale, we adopt $g_\star(T_i) = 106.75$.
For temperatures above the electroweak scale, we adopt the SM value $g_\star(T_i)=106.75$. More refined calculations including thermal corrections and the QCD crossover~\cite{Saikawa:2018rcs} produce only mild variations on $g_\star$ and do not significantly affect the qualitative behavior relevant for the runaway accretion phase considered here. Combining Eqs.~\eqref{eq:formation} and~\eqref{eq:rhor} yields the temperature of the primordial plasma at PBH formation time,
\begin{equation}
    T_i = \left( \frac{45 m_{\rm Pl}^2}{16 \pi^3  g_\star(T_i)} \frac{1}{t_i^2} \right)^{1/4} \,.
\end{equation}

\subsection{Accretion of primordial plasma}

During the radiation-dominated epoch the primordial plasma consists of relativistic SM particles in thermal equilibrium. PBHs embedded in this environment can accrete energy from the surrounding bath. The accretion rate may be estimated from the flux of particles gravitationally captured by the PBH. At temperature $T$, the phase-space distribution of the relativistic species $i$ with mass $m_i$, number of degrees of freedom $g_i$, and momentum $p$ is
\begin{equation}
    f_i(p)=\frac{1}{e^{E/T} \pm 1}\,,
\end{equation}
where the upper (lower) sign corresponds to Fermi--Dirac (Bose--Einstein) statistics and $E=E_i(p)=\sqrt{p^2+m_i^2}$. The total mass growth rate due to particle capture is then
\begin{equation}
    \label{eq:general_accretion}
    \frac{{\rm d}M_{\rm PBH}}{{\rm d}t} = \sum_i \int \frac{g_i\,\mathrm{d}^3p}{(2\pi)^3}\, f_i(p)\, E\, v\, \sigma_{\rm cap}(E)\,,
\end{equation}
where $v$ is the particle velocity and $\sigma_{\rm cap}(E)$ is the gravitational capture cross section. For relativistic particles incident on a Schwarzschild PBH, the capture cross section is determined by the critical impact parameter associated with unstable photon orbits~\cite{Misner:1973prb, Page:1976df}. Particles with impact parameter below this critical value are gravitationally captured, leading to
\begin{equation}
	\label{eq:capture_cross_section}
	\sigma_{\rm cap} = \pi\Big(\frac{3\sqrt{3}}{2}r_h\Big)^2 \,.
\end{equation}
In the relativistic limit, the cross section becomes effectively independent of particle momentum and can be factored out of the phase-space integral in Eq.~\eqref{eq:general_accretion}. In the ADD framework, SM particles, including the relativistic plasma responsible for early-time accretion, are confined to the standard 3-brane. Consequently, their dynamics are governed by the effective geometry induced on the brane rather than the full bulk volume. The gravitational capture cross section retains the geometric scaling $\sigma_{\rm cap} \propto \pi r_h^2$. The enhancement of the accretion rate in the ADD scenario is mainly driven by the larger horizon radius $r_h$ at fixed mass. For this reason, Eq.~\eqref{eq:capture_cross_section} gives a phenomenological description of the leading geometric enhancement expected in the ADD framework. In the relativistic limit, the remaining integral corresponds to the radiation energy density $\rho_r$, implying an accretion rate proportional to the horizon area and to the ambient radiation density.

This geometrical estimate corresponds to a ballistic regime, in which particles follow geodesic trajectories and are captured if their impact parameter is sufficiently small. However, its validity is limited in a dense and collisional plasma, where frequent particle interactions can disrupt ballistic trajectories. To assess the relevant regime, we compare the PBH horizon radius with the thermal mean free path of relativistic particles,
\begin{equation}
    \lambda_T \sim \frac{1}{\alpha_{\rm EM}^2 T}\,,
\end{equation}
where $\alpha_{\rm EM}\simeq 1/137$ is the electromagnetic fine-structure constant. Evaluating this ratio at the formation temperature $T_i$ gives
\begin{align}
    \frac{\lambda_T}{r_h} &\simeq \frac{M_\star}{a_n}\frac{1}{\alpha_{\rm EM}^2\,T_i}\left(\frac{M_\star}{M_{\rm PBH}}\right)^{\frac{1}{n+1}} \\
    & \simeq \frac{1}{a_n}\frac{1}{\alpha_{\rm EM}^2} \left( \frac{16 \pi^3  g_\star(T_i)M_\star^6} {45 \gamma^2 m_{\rm Pl}^6}\right)^{1/4}
 \left(\frac{M_{\rm PBH}}{M_\star}\right)^{\frac{1}{2} - \frac{1}{n+1}}\,,\nonumber
\end{align}
where we used Eq.~\eqref{eq:rh_piecewise} for the horizon radius in the higher-dimensional regime. For the parameter range of interest, one typically finds $\lambda_T \ll r_h$, implying that the plasma is strongly collisional near the PBH horizon. In this regime, particles undergo many scatterings before traversing the gravitational capture region, and the ballistic approximation breaks down. The accretion flow is then more appropriately described hydrodynamically, in terms of relativistic Bondi-like accretion. The resulting mass accretion rate can be expressed as~\cite{Nayak:2009wk}
\begin{equation}
    \label{eq:accretion}
    \dfrac{{\rm d}M_{\rm PBH}}{{\rm d}t}\bigg|_{\rm acc} = f_{\rm acc}\, 4\pi\, r_h^2\, \rho_r\,,
\end{equation}
where the dimensionless parameter $f_{\rm acc}$ characterizes deviations from the ideal Bondi rate~\cite{Bondi:1952ni}. Although the PBH geometry is $(4+n)$-dimensional for $r_h < R$, the ambient radiation bath is composed of SM degrees of freedom, which in the ADD framework is confined to the brane. Eq.~\eqref{eq:accretion} is an effective brane accretion rate controlled by the geometric capture cross section. All uncertainties associated with relativistic hydrodynamics, particle interactions, and the matching between the plasma and the near-horizon geometry are absorbed into the parameter $f_{\rm acc}$.

In a radiation-dominated background, the relativistic equation of state implies a large sound speed $c_s=1/\sqrt{3}$, which generally reduces the accretion efficiency onto PBHs. Accordingly, the literature typically favors efficiencies $f_{\rm acc}\ll1$, with representative values in the range $f_{\rm acc}\sim10^{-3}$--$10^{-1}$~\cite{Nayak:2009wk, Ye:2026ffe}. Early analyses already pointed out that accretion onto PBHs in the radiation-dominated era is typically inefficient, as the rapid cosmic expansion and relativistic pressure prevent the formation of a steady inflow~\cite{Niemeyer:1999ak}. More recent work has revisited this question using detailed numerical and semi-analytical approaches, emphasizing that the accretion efficiency is highly uncertain and depends sensitively on plasma effects, relative velocities, and radiative feedback~\cite{Jangra:2024sif}. In particular, Ref.~\cite{Jangra:2024sif} finds that cosmological accretion can deviate significantly from the simple Bondi prescription once the dynamical coupling between PBHs and the expanding background is properly included. In our numerical analysis, we adopt the conservative benchmark value $f_{\rm acc} = 10^{-3}$. Larger values would enhance the early growth phase and further strengthen our conclusions. Our results should be interpreted within this effective framework. Importantly, the conclusions of Refs.~\cite{Niemeyer:1999ak, Jangra:2024sif} are derived within standard four-dimensional gravity. In the ADD scenario, the enlarged horizon radius and suppressed Hawking temperature modify
%\textcolor{blue}{\sout{both the capture cross section and}}
the thermodynamic balance between accretion and evaporation.
%\textcolor{blue}{\sout{These effects}}
This effect can enhance the relative importance of accretion and may reopen a regime of efficient growth in the early Universe. A fully consistent treatment would require relativistic radiation hydrodynamics in a higher-dimensional background, which lies beyond the scope of the present work. This prescription should be interpreted as a phenomenological parameterization of accretion rather than a first principles calculation in a fully $(4+n)$-dimensional hydrodynamical framework.

\section{Revisiting cosmological evolution with accretion and backreaction}
\label{sec:evolution}

The cosmological evolution of PBHs in the ADD framework is governed by the coupled interplay between Hawking evaporation, radiation accretion, and cosmic expansion~\cite{Carr:1974nx, Carr:2017jsz}. A consistent treatment requires simultaneously tracking the evolution of the PBH mass, the radiation bath, and the Hubble expansion rate. Many previous studies of PBH growth assumed a fixed radiation background and neglected the dynamical backreaction of accretion and evaporation on the cosmological evolution~\cite{Bean:2002kx, Ricotti:2007au, Carr:2017jsz}. More recent analyses have emphasized that this approximation can fail when accretion becomes efficient or when PBHs constitute a non-negligible fraction of the energy density~\cite{Jangra:2024sif}. In contrast, we solve a fully coupled system of cosmological equations, consistently tracking PBH mass evolution, radiation, and expansion, including the dynamical backreaction of accretion and evaporation. This is particularly important in the ADD framework, where suppressed Hawking temperatures can render accretion highly efficient and qualitatively alter the evolution. The mass evolution of a PBH is determined by the competition between Hawking evaporation and radiation accretion. While the evaporation rate is given in Eq.~\eqref{eq:dMdt_evap}, accretion of radiation from the surrounding thermal bath is described by Eq.~\eqref{eq:accretion}. Using Eqs.~\eqref{eq:rhor} and~\eqref{eq:HawkingLED} yields the net mass evolution equation,
\begin{equation}
    \label{eq:dMdt_tot}
    \dfrac{{\rm d}M_{\rm PBH}}{{\rm d}t} = -\alpha(n)T_H^2 + \eta(n)\, \frac{T^4}{T_H^2}\,,
\end{equation}
where
\begin{equation}
    \eta(n) = \frac{\pi}{120}(n+1)^2 f_{\rm acc} g_\star \,.
\end{equation}
The description in Eq.~\eqref{eq:dMdt_tot} assumes a relativistic thermal bath with approximately constant $g_\star$, neglecting possible deviations from local thermal equilibrium near the PBH horizon. The inverse dependence on $T_H^2$ implies that accretion is strongly enhanced in the LED regime, where the Hawking temperature is suppressed, allowing it to dominate over evaporation even at relatively low plasma temperatures. As a result, Eq.~\eqref{eq:dMdt_tot} describes a continuous exchange of energy between the PBH population and the surrounding radiation bath. This expression is strictly valid in the higher-dimensional regime; in practice, it can be used as an effective interpolation across the transition, with $T_H$ and $r_h$ defined by the piecewise prescription in Eq.~\eqref{eq:rh_piecewise}.

The coupled evolution equations for the radiation and PBH energy densities are
\begin{align}
    \label{eq:density_evolution}
    \dfrac{{\rm d}\rho_r}{{\rm d}t} + 4H\rho_r &= -\Gamma \rho_{\rm PBH} \,, \\
    \dfrac{{\rm d}\rho_{\rm PBH}}{{\rm d}t} + 3H\rho_{\rm PBH} &= \Gamma \rho_{\rm PBH} \,,
\end{align}
where $\Gamma \equiv \dot{M}_{\rm PBH}/M_{\rm PBH}$ represents the fractional PBH mass growth rate resulting from the combined effects of accretion and evaporation. The cosmological expansion is determined by the total energy density through the Friedmann equation
\begin{equation}
    H^2 = \frac{8\pi}{3 m_{\rm Pl}^2}\,\left(\rho_r+ \rho_{\rm PBH}\right)\,.    
\end{equation}
Including the PBH contribution to the expansion rate is essential in the LED framework. Because accretion can rapidly increase the PBH mass, a significant fraction of the radiation energy density may be transferred to the PBH population over a relatively short cosmological time. Neglecting this backreaction would break the self-consistency between mass growth, energy conservation, and cosmic expansion. This formulation assumes conservation of the comoving PBH number density after formation, so that the evolution of $\rho_{\rm PBH}$ is entirely driven by mass growth and cosmic expansion.

To characterize the competition between evaporation and accretion, we introduce the balance temperature $T_{\rm bl}$, defined as the plasma temperature at which the two processes exactly compensate each other in Eq.~\eqref{eq:dMdt_tot}. Solving Eq.~\eqref{eq:dMdt_tot} for ${\rm d}M_{\rm PBH}/{\rm d}t = 0$ yields
\begin{equation}
    T_{\rm bl} = \left( \frac{\alpha(n)}{\eta(n)} \right)^{1/4} \frac{n+1}{4\pi} \frac{M_\star}{a_n} \left( \frac{M_\star}{M_{\rm PBH}} \right)^{1/(n+1)}\,.
\end{equation}
When the plasma temperature satisfies $T>T_{\rm bl}$, accretion dominates over Hawking evaporation and the PBH mass increases. Conversely, for $T<T_{\rm bl}$ evaporation becomes the dominant process. Table~\ref{tab:thresholds} shows the strong dependence of this threshold on the number of extra dimensions. In the standard four-dimensional case $n=0$, the temperature required for accretion to overcome evaporation is higher. In contrast, the LED scenario significantly lowers the threshold, allowing accretion to dominate under realistic early-Universe conditions.

\begin{table}[htb]
\centering
\renewcommand{\arraystretch}{1.4}
\setlength{\tabcolsep}{12pt}
\begin{tabular}{lccc}
\hline
Model & $n$ & $T_H$ [GeV] & $T_{\rm bl}$ [GeV] \\
\hline
Standard 4D & 0 & $\sim 10\phantom{^-0}$ & $\sim 10^2\phantom{0}$ \\
LED & 2 & $\sim 10^{-8}$  & $\sim 10^{-7}$  \\
LED & 4 & $\sim 10^{-3}$  & $\sim 10^{-2}$  \\
LED & 6 & $\sim 10^{-1}$  & $\sim 1\phantom{0^-1}$  \\
\hline
\end{tabular}
\caption{Comparison of the balance temperature $T_{\rm bl}$ for a PBH of mass $M_{\rm PBH}=10^{12}$\,g (in the higher-dimensional regime), assuming $M_\star=10$\,TeV and $f_{\rm acc}=10^{-3}$.}
\label{tab:thresholds}
\end{table}

The evolution of the plasma temperature is obtained from Eq.~\eqref{eq:density_evolution},
\begin{equation}
\label{eq:dT_dt}
    \frac{{\rm d}T}{{\rm d}t} = -H T - \frac{\Gamma}{4} \frac{\rho_{\rm PBH}}{\rho_r} T \,,
\end{equation}
where the first term describes the standard adiabatic cooling due to cosmic expansion, while the second term accounts for the energy exchange between PBHs and radiation. Depending on the sign of $\Gamma$, this term represents either the depletion of radiation during the accretion phase or the injection of thermal energy via Hawking evaporation. The initial conditions are set at the PBH formation time $t_i$ as derived in Sec.~\ref{sec:initial_conditions},
\begin{equation}
    \begin{cases}
        t_i = \dfrac{M_i}{\gamma m_{\rm Pl}^2}, \\[10pt]
        T_{i} = \left( \dfrac{45 m_{\rm Pl}^2}{16 \pi^3 g_\star} \dfrac{1}{t_i^2} \right)^{1/4}, \\[10pt]
        \rho_{\rm PBH}^i = \dfrac{\beta}{1-\beta}\rho_{r}(T_i) \approx \beta\,\rho_{r}(T_i);
    \end{cases}
\end{equation}
where $\beta$ is the initial mass fraction of the Universe collapsing into PBHs at the formation time $t_i$. For PBHs forming during the radiation-dominated era, the formation time is directly related to the horizon mass. For example, PBHs with an initial mass $M_i \sim 10^{15}$\,g form at $t_i \sim 10^{-23}$\,s. These masses are many orders of magnitude above the fundamental gravity scale $M_\star$, ensuring that the PBHs are safely in the semiclassical regime and do not evaporate immediately. At the same time, they remain well below the transition mass to the four-dimensional regime, $M_i \ll M_{\rm 4D}$, so the PBHs still reside in the higher-dimensional regime. In this regime, the Hawking temperature is strongly suppressed, providing ideal conditions for efficient accretion from the surrounding primordial plasma.

\begin{figure}[htb]
    \centering
    \includegraphics[width=0.45\textwidth]{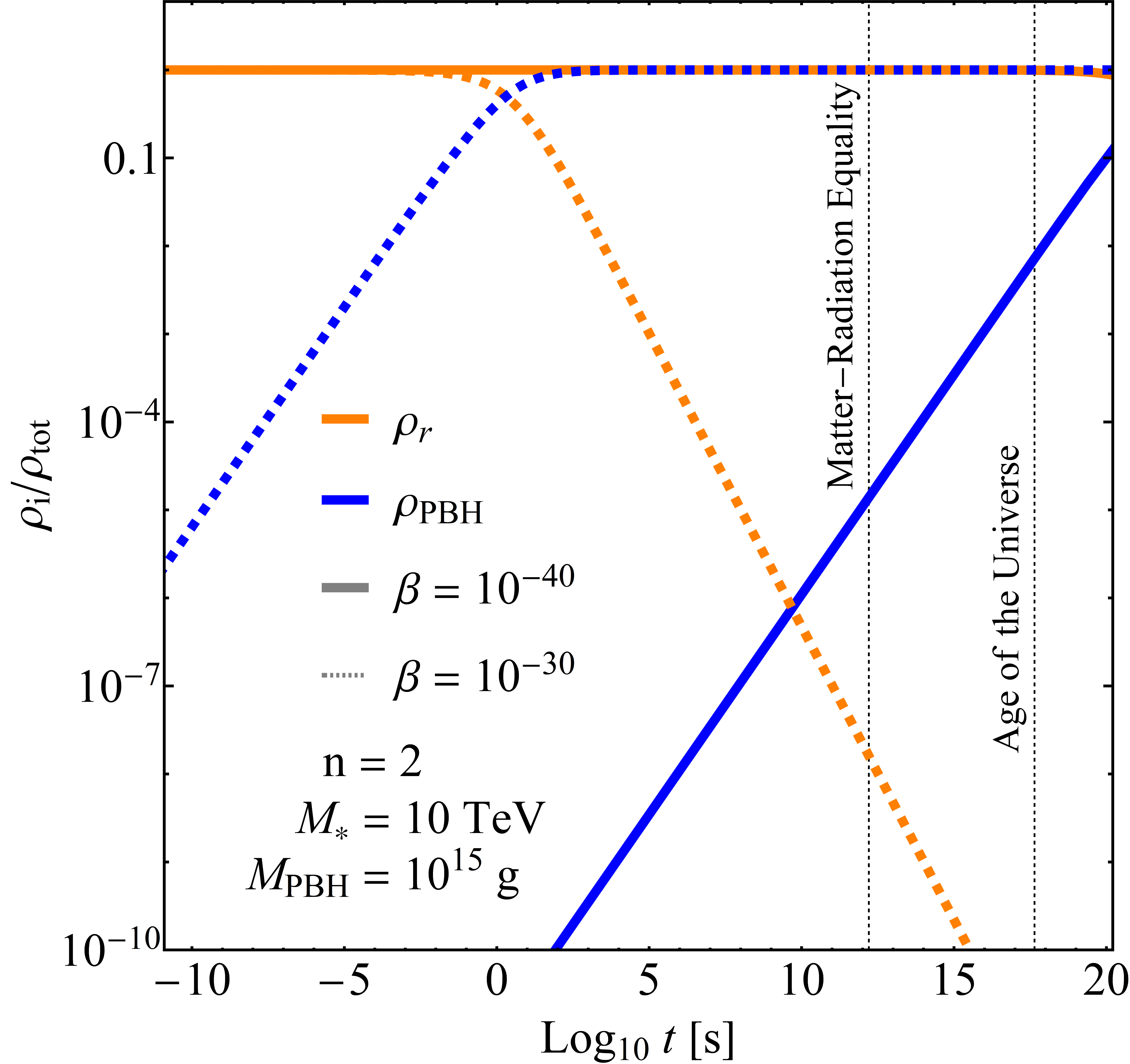} 
    \caption{Evolution of the energy density ratios $\rho_i/\rho_{\rm tot}$ for radiation (orange) and PBHs (blue) as a function of time for $n=2$ extra dimensions. The solid lines correspond to an initial abundance $\beta = 10^{-40}$, while the dashed lines show $\beta = 10^{-30}$.}
    \label{fig:rho_evolution}
\end{figure}

With these equations and initial conditions, the full cosmological system can be solved numerically to track the coupled evolution of the PBH population and the radiation bath. Figure~\ref{fig:rho_evolution} shows the evolution of the energy density fractions $\rho_r/\rho_{\rm tot}$ and $\rho_{\rm PBH}/\rho_{\rm tot}$, where $\rho_{\rm tot} = \rho_r+ \rho_{\rm PBH}$, as a function of cosmic time. The evolution is tracked from the PBH formation epoch through the matter-radiation equality time $t_{\rm eq} \sim 10^{12}$\,s, extending up to and beyond the present age of the Universe $t_0 \sim 10^{17}$\,s. The numerical results reveal that, for models with extra dimensions $n \geq 2$, the early accretion phase can dramatically increase the PBH mass shortly after formation. As a consequence, the PBH energy density grows more rapidly than expected for a fixed-mass non-relativistic component. Even extremely small initial abundances $\beta$ can evolve into a significant DM component in PBHs at late times.

This behavior originates from a phase of runaway accretion immediately following PBH formation. In the $(4+n)$-dimensional regime, the Hawking temperature is suppressed, while the primordial plasma temperature remains very high. The resulting hierarchy $T \gg T_H$ leads to an accretion rate $\dot{M}_{\rm acc} \propto T^4/T_H^2$ that is extremely efficient. Consequently, the PBH mass can increase by several orders of magnitude within a fraction of a Hubble time after formation. 

At early times the Universe is radiation-dominated so that $\rho_r/\rho_{\mathrm{tot}}\simeq 1$ (orange solid and dashed lines) and $\rho_{\mathrm{PBH}}/\rho_{\mathrm{tot}}\ll 1$ (blue solid and dashed lines). In the case with larger initial abundance, $\beta=10^{-30}$, the PBH fraction grows rapidly and soon approaches unity (blue-dashed). This behavior reflects the efficient early-time accretion in the higher dimensional regime, where the Hawking temperature is suppressed and the surrounding radiation bath remains sufficiently hot to drive net mass growth. Therefore, PBHs quickly dominate the total energy density and the radiation fraction correspondingly decreases (orange dashed).

For the much smaller initial abundance $\beta=10^{-40}$, the PBH contribution remains negligible for a much longer period, and the Universe stays radiation-dominated over most of the evolution shown (orange solid). Nevertheless, the PBH fraction steadily increases with time (blue solid). The late time growth is due to the fact that PBHs dilute as non-relativistic matter whereas the radiation redshifts faster.

\begin{figure}[htb]
    \centering
    \includegraphics[width=0.42\textwidth]{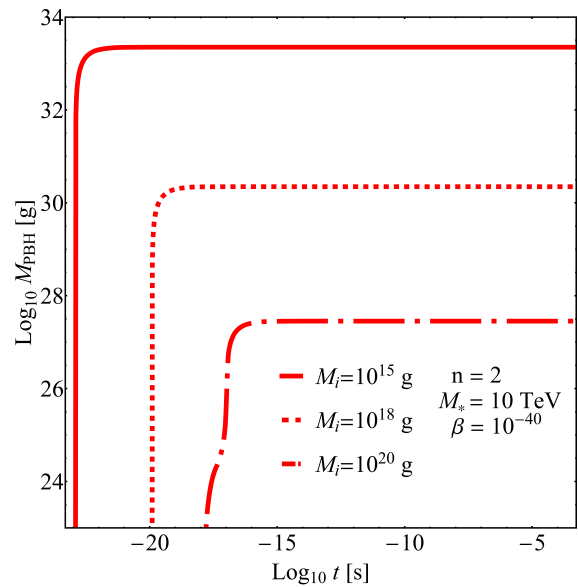}
    \caption{Evolution of the PBH mass for $n=2$ extra dimensions and an initial abundance $\beta = 10^{-40}$. The curves correspond to initial masses $M_i = \{10^{15}, 10^{18}, 10^{20}\}$\,g.}
    \label{fig:mass_evolution}
\end{figure}

The mass evolution during this early phase is illustrated in Fig.~\ref{fig:mass_evolution}, where trajectories corresponding to different initial PBH masses are shown. For the case $n=2$, the critical transition mass is $M_{\rm 4D} \sim 10^{24}$\,g. This behavior is driven by an early phase of efficient accretion in the higher-dimensional regime. During this phase the PBH mass rapidly approaches $M_{\rm 4D}$, at which point the horizon radius becomes comparable to the compactification scale. The lightest initial seeds grow the most (red solid line), since they form earlier and therefore remain longer in the high temperature environment where accretion is most efficient. Once this transition occurs, the system enters the standard four-dimensional Schwarzschild regime. The Hawking temperature then follows the usual inverse-mass scaling, and the accretion rate evolves more gradually. Despite this transition, the early runaway accretion phase can already increase the PBH mass by many orders of magnitude, allowing initially microscopic seeds to reach macroscopic masses.

\section{Results}
\label{sec:results}

We now determine the critical initial abundance $\beta_{\rm crit}$, which sets the PBH energy density at formation. Our goal is to identify the value of $\beta_{\rm crit}$ required for PBHs to account for the entire DM energy density at matter-radiation equality. For the numerical analysis, we restrict the scan to initial masses $M_i \ge 10^{12}\,$g, for which the transition from the higher-dimensional regime to the subsequent four-dimensional evolution can be robustly resolved. In this regime, PBHs may initially experience enhanced accretion, but once their horizon radius exceeds the compactification scale, their evolution becomes effectively four-dimensional and accretion rapidly becomes inefficient. This ensures that the late-time evolution is numerically stable and well described by standard Schwarzschild dynamics. This also clarifies that exploring smaller $M_\star$ or larger $n$ increases both the duration and efficiency of the early higher-dimensional accretion phase. At matter-radiation equality, $\rho_m(t_{\rm eq}) \simeq \rho_r(t_{\rm eq})$, and the DM energy density at that epoch is
\begin{equation}
    \rho_{\rm DM}(t_{\rm eq}) = \frac{\Omega_{\rm DM}}{\Omega_m}\,\rho_r(t_{\rm eq})\,,
\end{equation}
where we set $\Omega_{\rm DM}=0.264$ and $\Omega_m = 0.313$~\cite{Planck:2018vyg}. We determine $\beta_{\rm crit}$ by requiring that the PBH density at $t_{\rm eq}$ saturates the DM abundance,
\begin{equation}
    f_{\rm{PBH}} \equiv \frac{\rho_{\rm PBH}(t_{\rm eq})}{\left( \frac{\Omega_{\rm DM}}{\Omega_m} \right) \rho_r(t_{\rm eq})} = 1\,.
\end{equation}

\begin{figure}[htb]
    \centering
    \includegraphics[width=0.45\textwidth]{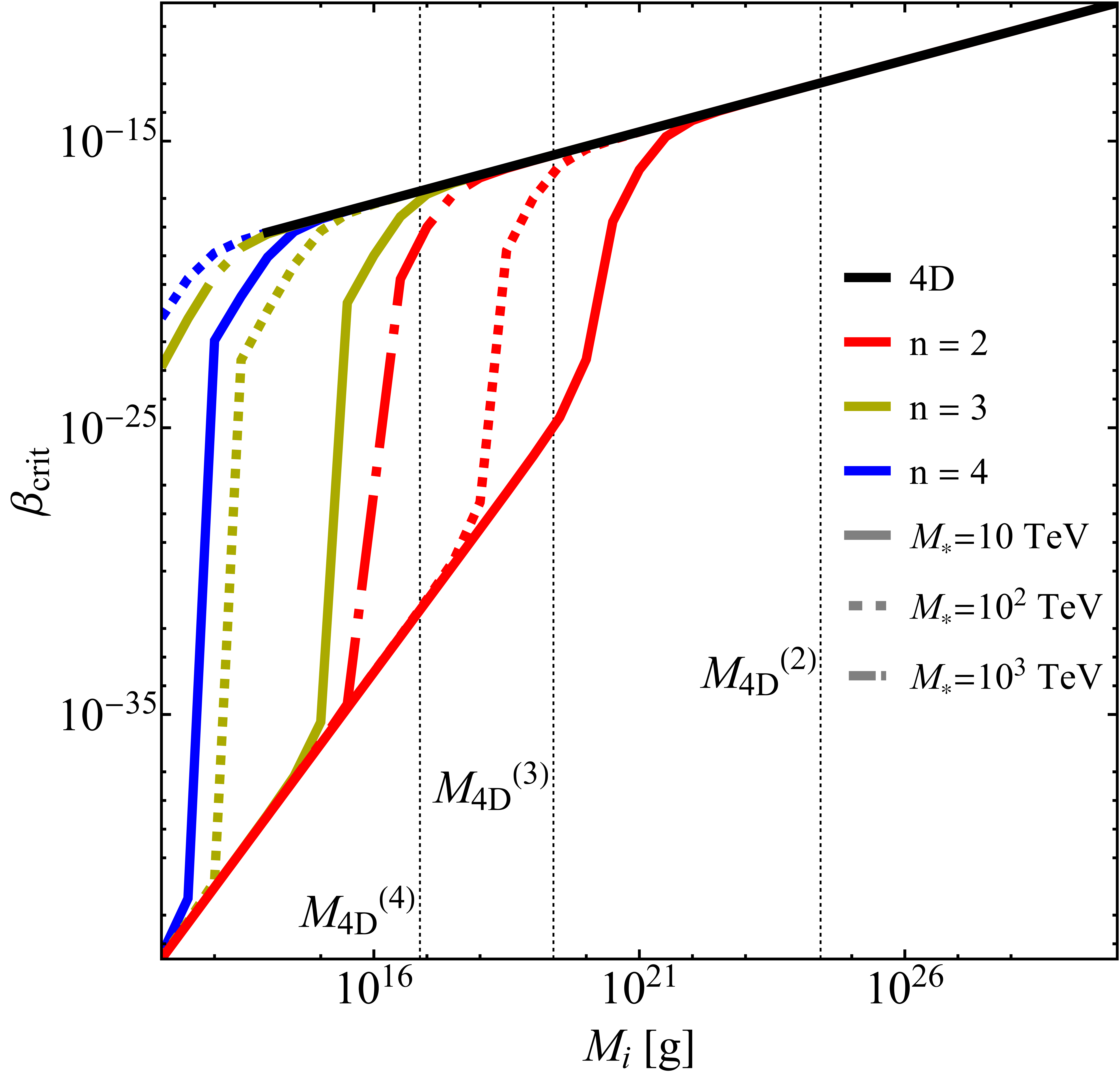}
    \caption{Critical initial abundance $\beta_{\rm crit}$ required for PBHs to constitute the entirety of DM ($f_{\rm PBH}=1$) at matter–radiation equality as a function of the initial mass $M_i$. The standard 4D scenario (black solid line) follows the expected scaling $\beta \propto \sqrt{M_i}$. Higher-dimensional cases are shown for $n=2$ (red), $n=3$ (yellow), and $n=4$ (blue). For each $n$, different fundamental gravity scales are considered: $M_\star = 10$\,TeV (solid), $10^2$\,TeV (dashed), and $10^3$\,TeV (dotted). Vertical dashed lines indicate the transition mass $M_{\rm 4D}$ for given values of $n$, beyond which PBHs form directly in the four-dimensional regime.}
    \label{fig:beta_critic}
\end{figure}

In Fig.~\ref{fig:beta_critic} we show $\beta_{\rm crit}$ as a function of the initial PBH mass $M_i$ for different numbers of extra dimensions, together with the standard four-dimensional result. The presence of LED significantly enlarges the parameter space in which PBHs can account for the total DM density, as the early higher-dimensional accretion phase amplifies the PBH mass shortly after formation. Consequently, much smaller initial abundances are required to reproduce the observed DM density. For instance, in the case $n=2$ with $M_\star = 10$\,TeV, an initial fraction as small as $\beta \sim 10^{-44}$ at $M_i \sim 10^{12}$\,g is sufficient to achieve $f_{\rm PBH}=1$. This strong suppression of $\beta$ reflects the large mass amplification factor accumulated during the early growth phase. Part of the required tuning is effectively shifted from the initial abundance to the efficiency of early-Universe accretion. We emphasize that these results are obtained under the assumption of a monochromatic PBH mass function. In realistic formation scenarios, extended mass distributions could modify both the accretion history and the mapping to the DM abundance. Finally, the extremely small values of $\beta$ correspond to rare primordial fluctuations in the high-density tail of the distribution. In Gaussian scenarios, $\beta$ depends exponentially on the variance of density perturbations, so that even very small collapse fractions can arise from modest changes in the primordial power spectrum.

The transition to the standard four-dimensional behavior is also clearly visible in Fig.~\ref{fig:beta_critic}. As the initial mass $M_i$ approaches the transition mass $M_{\rm 4D}$, defined by $r_h \sim R$, the influence of the higher-dimensional phase gradually disappears and the curves converge toward the standard cosmological trajectory. Physically, when $M_i \gtrsim M_{\rm 4D}$ the PBH horizon radius already exceeds the compactification scale, and the PBH therefore evolves entirely within a four-dimensional spacetime. Beyond these critical masses, the effects of the higher-dimensional phase become negligible, and the system follows the standard cosmological behavior. In this regime the required abundance follows the relation
\begin{equation}
    \beta_{\rm{crit}} \approx \frac{\Omega_{\rm DM}}{\Omega_m}\left( \frac{T_{\rm{eq}}}{T_i}\right) \approx \frac{\Omega_{\rm DM}}{\Omega_m} \left( \frac{16 \pi^3 g_\star }{45 \gamma^2 m_{\rm Pl}^6} \right)^{1/4}  T_{\rm{eq}} \sqrt{M_i}\,,
    \label{eq:beta_asymptotic}
\end{equation}
where $T_{\rm{eq}}$ is the temperature at matter-radiation equality and $T_i$ is the temperature of the plasma at PBH formation. This behavior yields the characteristic scaling $\beta_{\rm crit}\propto \sqrt{M_i}$ and follows directly from the relation between the horizon mass and the plasma temperature, $M_i=\gamma M_H(T_i)\propto T_i^{-2}$. As shown in Fig.~\ref{fig:beta_critic}, all higher-dimensional solutions asymptotically converge toward this four-dimensional limit once the early accretion phase is absent. The recovery of the standard scaling therefore provides an important consistency check of the numerical implementation.

\begin{figure}[htb]
    \centering
    \includegraphics[width=0.45\textwidth]{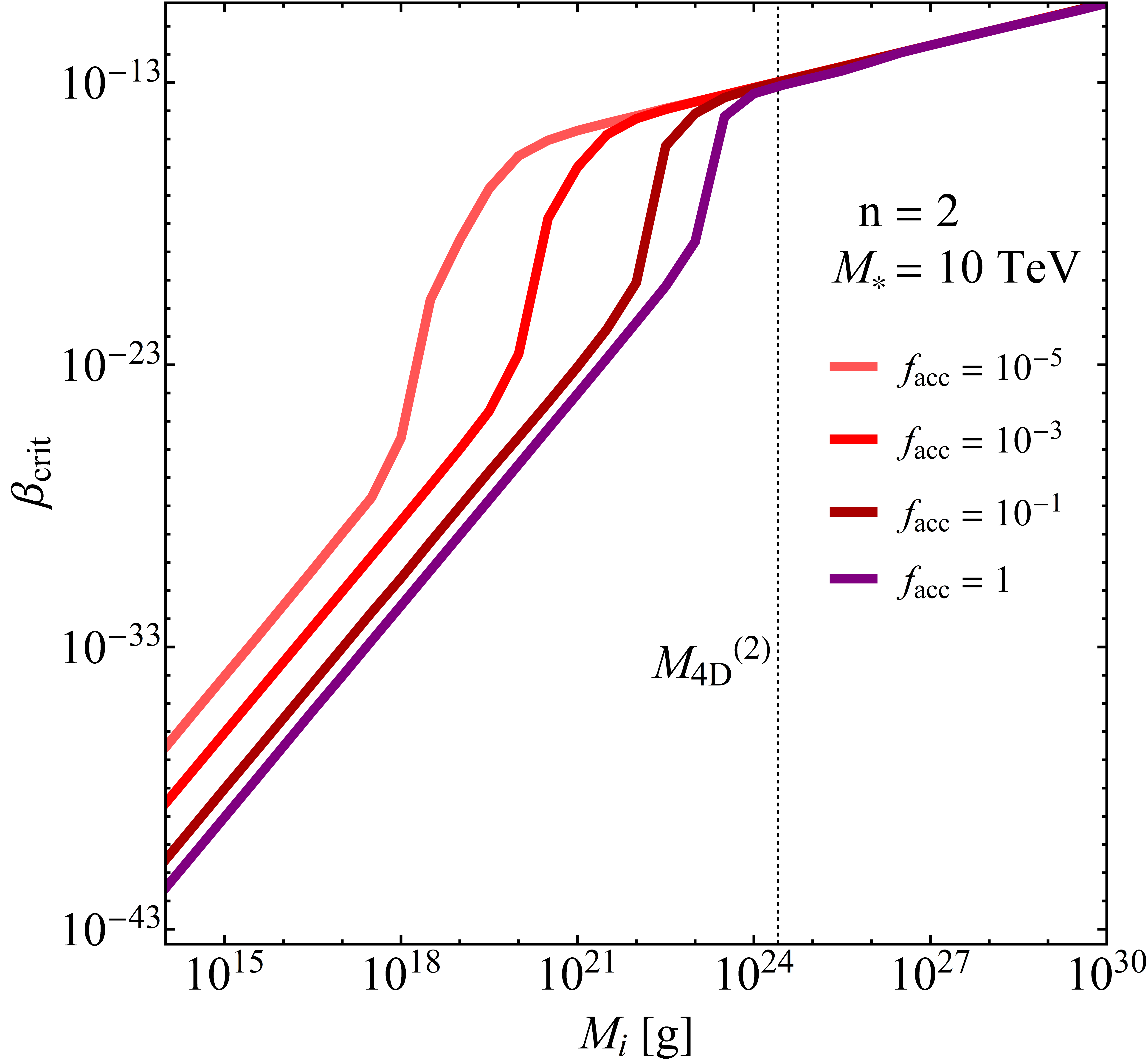}
    \caption{Impact of the accretion efficiency $f_{\rm acc}$ on the critical collapse fraction $\beta_{\rm crit}$ for $n=2$ and $M_\star = 10$\,TeV. Color shades from light red to purple represent increasing values of $f_{\rm acc}$ from $10^{-5}$ to $1$. The plot illustrates the degeneracy between the initial abundance and the efficiency of extra-dimensional accretion processes. }
    \label{fig:beta_criticvsf_acc}
\end{figure}

In Fig.~\ref{fig:beta_criticvsf_acc} we perform a dedicated sensitivity analysis for the $n=2$ case, exploring how the accretion efficiency $f_{\rm acc}$ affects the required initial abundance. As $f_{\rm acc}$ decreases from $1$ to $10^{-5}$, the critical abundance $\beta_{\rm crit}$ increases by several orders of magnitude. This nearly-linear scaling in the ADD regime highlights a fundamental degeneracy, the current DM abundance can be explained either by a very small population of highly efficient accreting PBHs or by a slightly larger population with lower accretion rates. Crucially, the convergence of all curves toward the standard 4D trajectory for $M_i \gtrsim M_{\rm 4D}$ for $n=2$ demonstrates that the impact of $f_{\rm acc}$ is strictly confined to the higher-dimensional phase, ensuring the consistency of the model with late-time standard cosmology. We now turn to the implications of these results in light of observational constraints.

\section{Discussion}
\label{sec:discussion}

A key outcome of our analysis is the strong sensitivity of PBH cosmology to the presence of large extra dimensions. The runaway accretion mechanism significantly reshapes the allowed parameter space for the initial collapse fraction $\beta$ by remapping late-time astrophysical constraints onto much smaller formation mass scales. In practice, this remapping arises from the non-trivial relation between $M_i$ and $M_f$ induced by early accretion. In particular, microlensing limits from the Exp\'erience pour la Recherche d'Objets Sombres (EROS-2)~\cite{EROS-2:2006ryy}, the Optical Gravitational Lensing Experiment (OGLE-III)~\cite{Wyrzykowski:2011tr, Wyrzykowski:2015ppa} and subsequent release (OGLE-IV)~\cite{2015AcA....65....1U, Mroz:2024mse}, and the Hyper Suprime-Cam survey (HSC)~\cite{Niikura:2017zjd, Niikura:2019kqi} must be reinterpreted once the accretion-induced mapping between the initial PBH mass $M_i$ and the present-day mass $M_f$ is taken into account. To quantify this mapping, we relate the collapse fraction at formation to the present-day PBH abundance as follows.

To evaluate the viability of PBHs as DM candidates in ADD scenarios, we must reconcile early-Universe formation conditions with late-time astrophysical observations. Current constraints from microlensing surveys provide upper bounds on the current DM fraction, $f_{\rm max}(M_f)$, as a function of the observed mass $M_f$. While in standard 4D cosmology the PBH mass remains nearly constant ($M_i \simeq M_f$), in ADD models the early radiation-era accretion phase can produce a final mass $M_f$ that is orders of magnitude larger than the initial mass $M_i$, as shown in Fig.~\ref{fig:mass_evolution}. The mapping between the formation epoch and the present day relies on the following chain of relations. At the formation time $t_i$, the collapse fraction is defined as $\beta \approx n_i M_i / \rho_{\rm tot}(t_i)$. Under the assumption of a conserved comoving number density ($n_0 \propto n_ia^{-3}$), the present-day energy density of PBHs is given by $\rho_{PBH}(t_0) = n_0 M_f$, where $M_f$ accounts for the ADD-induced accretion. Since the DM density scales as $\rho_{DM} \propto a^{-3}$, the current fraction $f_{\rm PBH} = \rho_{\rm PBH}(t_0)/\rho_{DM}(t_0)$ is directly proportional to the initial $\beta$ scaled by the mass growth factor. At the level of parametric scaling, and neglecting changes in the PBH number density after formation, one has
\begin{equation}
    f_{\rm PBH}(M_f) \propto \beta(M_i) \frac{M_f}{M_i}\,,
\end{equation}
up to factors accounting for the precise evolution of the scale factor and the radiation-to-matter transition. This approximation captures the dominant effect of mass growth during the early accretion phase. This relation allows any observational upper bound $f_{\rm max}(M_f)$ to be translated into a constraint on the primordial collapse fraction $\beta_{\rm max}(M_i)$. This mapping assumes a conserved comoving PBH number density after formation, negligible merger-driven number evolution, and no significant late-time mass loss after the early accretion phase.

Specifically, by defining $\beta_{\rm crit}(M_i)$ as the collapse fraction required for PBHs to account for the full DM abundance at matter-radiation equality, we can determine the maximum allowed collapse fraction through
\begin{equation}
    \beta_{\rm max}(M_i) = \beta_{\rm crit}(M_i) \cdot f_{\rm max}(M_f(M_i))\,,
\end{equation}
where $M_f(M_i)$ is the numerical mapping obtained from the ADD accretion evolution.

The non-trivial relation between $M_i$ and $M_f$ induced by early-Universe accretion implies that PBHs formed with small initial masses undergo an efficient phase of runaway growth in the higher-dimensional regime, during which their masses can increase by several orders of magnitude. As a result, PBHs that initially lie far below observationally accessible mass ranges can evolve into the windows probed by microlensing surveys. While this mechanism eases the requirements for generating macroscopic PBHs from microscopic seeds, it simultaneously imposes strong restrictions on the PBH parameter space. Additional constraints arise from energy injection due to Hawking evaporation, which can affect the ionization history of the intergalactic medium and provide complementary bounds on the PBH abundance~\cite{Yin:2026hvw}.

Microlensing surveys constrain the DM fraction in compact objects by monitoring large stellar populations and searching for transient magnification events. The characteristic duration of these events is set by the Einstein timescale, $t_E \propto \sqrt{M_f}$, which determines the mass range to which each survey is most sensitive. The EROS-2 and OGLE-III/OGLE-IV surveys probe long-duration events associated with relatively heavy compact objects, typically in the range $M_f \sim 10^{26}$--$10^{34}$\,g, using wide-field monitoring of stars in the Magellanic Clouds and the Galactic bulge. In contrast, the HSC survey extends sensitivity to much shorter timescales and lower masses, $M_f \sim 10^{23}$--$10^{28}$\,g, through high-cadence observations of stars in M31~\cite{Niikura:2017zjd, Niikura:2019kqi}.

\begin{figure*}[t]
    \centering
    \subfigure{
        \includegraphics[height=7.5cm]{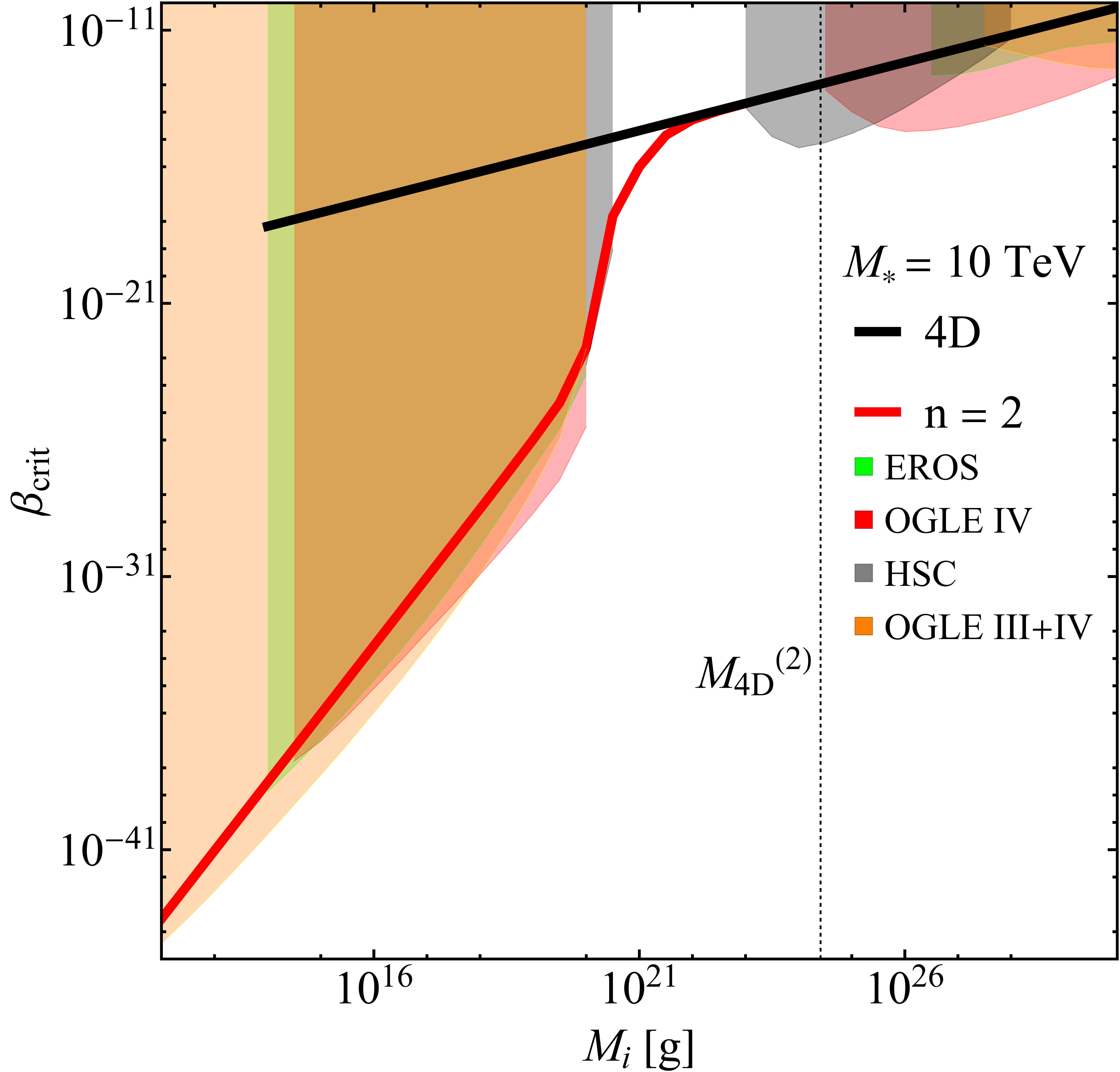}
    }\hfill
    \subfigure{
        \includegraphics[height=7.5cm,trim=1 0 0 0,clip]{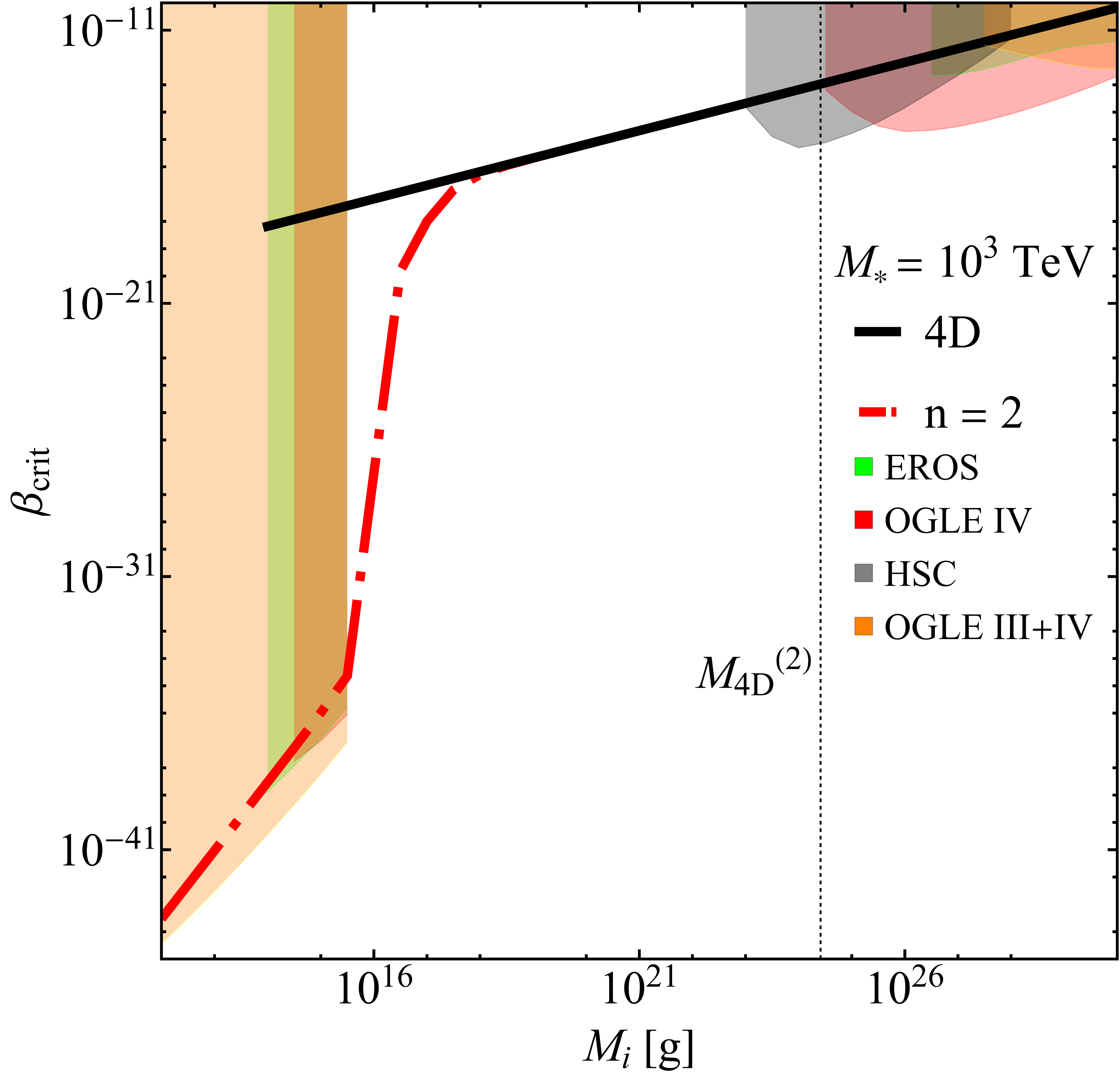}
    }\\[2mm]
    \subfigure{
        \includegraphics[height=7.5cm]{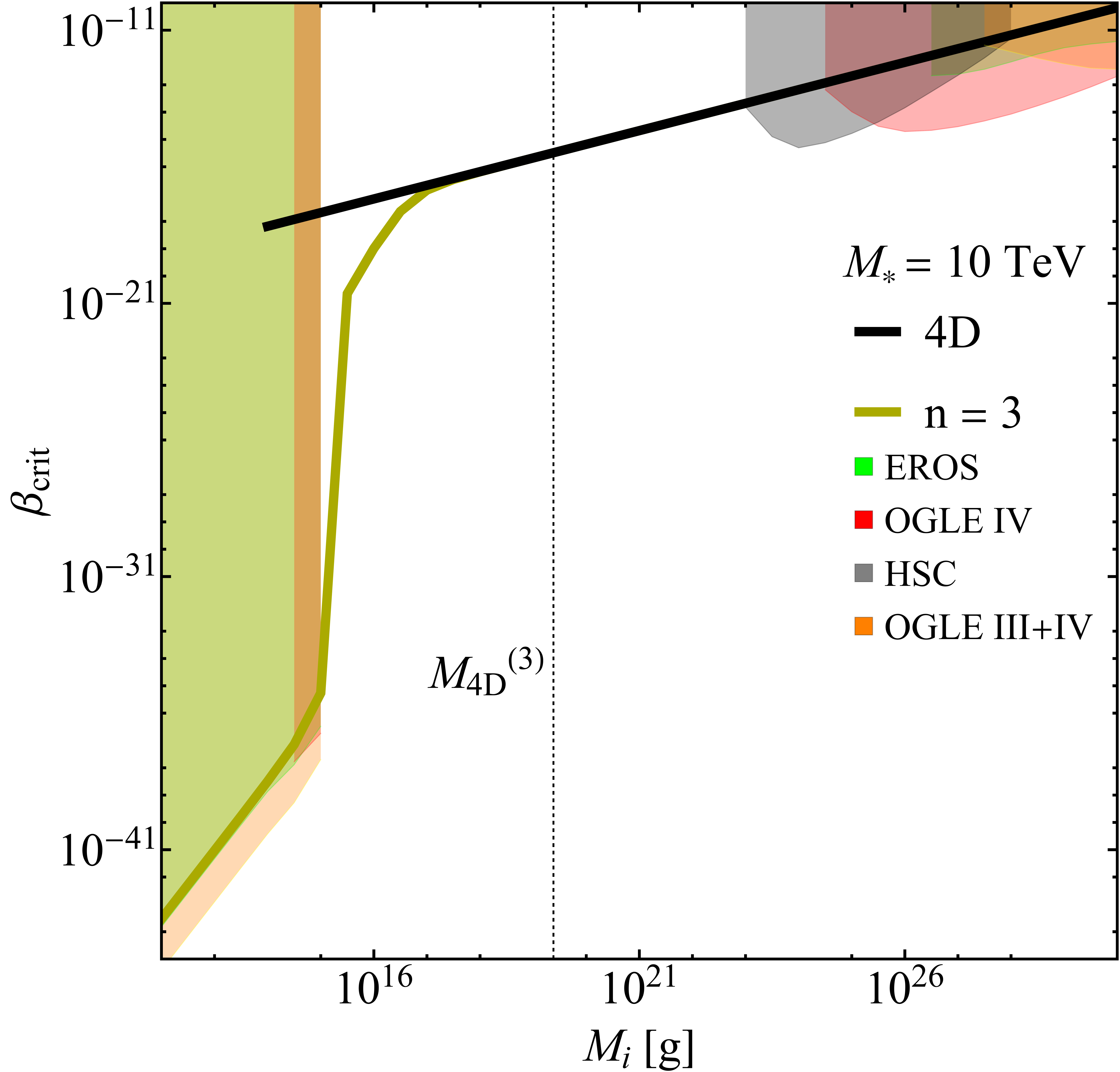}
    }\hfill
    \subfigure{
        \includegraphics[height=7.5cm,trim=1 0 0 0,clip]{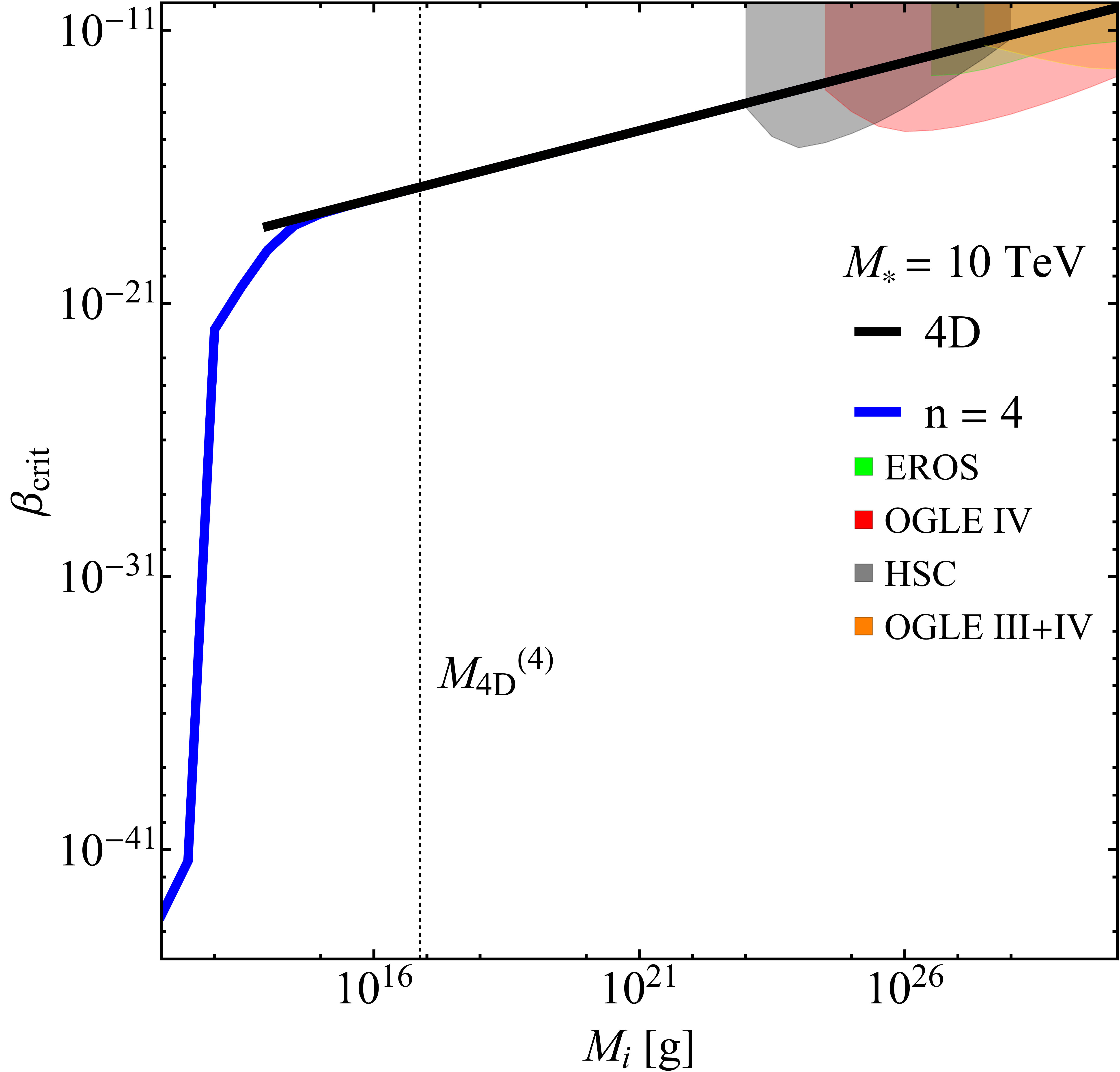}
    }
    \caption{Constraints on the primordial collapse fraction as a function of the formation mass $M_i$, including the effects of accretion in ADD scenarios. The solid black line denotes the standard 4D result, while colored curves show the critical abundance $\beta_{\rm crit}$ for extra-dimensional models with $n=2$ (red solid), $n=3$ (green solid), and $n=4$ (blue solid) for $M_\star = 10$\,TeV; red dot-dashed line corresponds to $M_\star = 10^3$\,TeV. Shaded regions indicate exclusion limits from microlensing surveys, EROS-2 (green), OGLE-III and OGLE-III/OGLE-IV (red/orange), and HSC (gray), translated into constraints on the initial collapse fraction $\beta_{\rm max}(M_i)$ using the accretion-induced mapping $M_i \rightarrow M_f$. Vertical dashed lines mark the transition mass $M_{\rm 4D}$ above which PBHs evolve in the standard four-dimensional regime.
    All panels assume a benchmark accretion efficiency of $f_{\rm acc} = 10^{-3}$; the impact of varying $f_{\rm acc}$ is detailed in Appendix~\ref{app:facc_dependence}.}
    \label{fig:beta_cnstrnt}
\end{figure*}

In the ADD framework, PBHs formed with masses well below these observational ranges can grow into them due to the early runaway accretion phase. Consequently, microlensing constraints derived at a given observed mass $M_f$ effectively exclude an extended range of initial masses $M_i$. The impact of this remapping is illustrated in Fig.~\ref{fig:beta_cnstrnt}, where observational limits are translated into bounds on the primordial collapse fraction as a function of $M_i$. As shown, the accelerated accretion characteristic of extra-dimensional models shifts microlensing constraints toward significantly smaller initial masses, approximately in the range $M_i \sim 10^{12}$--$10^{20}$\,g. PBHs formed in this interval grow efficiently and are mapped into mass ranges already constrained by microlensing observations, leading to their indirect exclusion and a substantial compression of the viable parameter space.

For the representative case $n=2$, the region in which PBHs can account for the total DM abundance is restricted to $M_i \approx 10^{21}$--$10^{22}$\,g (red solid). Smaller initial masses are excluded because their subsequent growth produces PBHs in observationally constrained mass ranges. Increasing the fundamental scale $M_\star$ suppresses early accretion by lowering the transition mass $M_{\rm 4D}$, shortening the higher-dimensional growth phase. This reduces the overall mass amplification and enlarges the viable parameter space, shifting the allowed region down to $M_i \approx 10^{17}$\,g (red dot-dashed). A similar qualitative behavior is observed for larger numbers of extra dimensions. For $n=3$ and $n=4$, the higher Hawking temperature associated with a smaller horizon radius reduces the net accretion rate, resulting in a weaker mass amplification. Consequently, the allowed regions occur at smaller initial masses, around $M_i \approx 10^{15}$\,g for $n=3$ (green solid) and $M_i \approx 10^{12}$\,g for $n=4$ (blue solid), where PBHs remain sufficiently light to evade current observational constraints. The black solid line represents the usual 4D PBH constraint, where the PBH mass does not undergo the enhanced early-time growth characteristic of the ADD scenario. Its lower-mass endpoint is set by the requirement that the PBH survive Hawking evaporation until the present epoch.

The constraints presented in Fig.~\ref{fig:beta_cnstrnt} are evaluated assuming the benchmark accretion efficiency $f_{\rm acc} = 10^{-3}$. The corresponding constraints are reported in Appendix~\ref{app:facc_dependence} for $f_{\rm acc} = 10^{-7}$, see Fig.~\ref{fig:beta_cnstrnt1em7}, and $f_{\rm acc} = 10^{-1}$, see Fig.~\ref{fig:beta_cnstrnt1em1}. These results illustrate how the mapping $M_i\rightarrow M_f$ varies with the phenomenological parameter $f_{\rm acc}$. Reducing the accretion efficiency suppresses the early runaway growth phase and widens the viable parameter windows in which PBHs can account for the total DM abundance.

While this mechanism provides a natural pathway for microscopic PBHs to grow into macroscopic DM candidates, it simultaneously imposes strong restrictions on the allowed parameter space. In particular, for $M_\star \sim \mathrm{TeV}$, a large region of small initial masses is effectively excluded unless the initial collapse fraction is extremely suppressed. This places non-trivial constraints on the underlying formation mechanism and provides a clear target for inflationary model building in the context of large extra dimensions, where compatibility with current observations may require a sharply peaked primordial power spectrum~\cite{Germani:2018jgr}. More generally, a complete assessment of the formation side of the scenario will require connecting the values of $\beta(M_i)$ identified here to the statistics of primordial curvature perturbations. 

Furthermore, microlensing constraints depend not only on the PBH mass, but also on the assumed Galactic halo model, velocity distributions, and survey-specific detection efficiencies. The expected event rate involves a convolution of the PBH mass function with the efficiency function of each experiment, which differs significantly between EROS-2, OGLE-III, OGLE-III/OGLE-IV, and HSC due to their distinct observational strategies. While our analysis captures the leading effect of mass growth through the mapping $M_i \rightarrow M_f$, a fully consistent comparison with observational data would require incorporating these efficiencies together with extended PBH mass functions. This is particularly relevant in the present scenario, where accretion can broaden the effective mass distribution even if the initial spectrum is monochromatic.

Our analysis assumes a monochromatic PBH mass function. Extending the framework to realistic broad mass distributions is an important direction for future work, since both the accretion history and the mapping to observational constraints may be modified. Likewise, a fully self-consistent treatment of accretion in the higher-dimensional regime would require relativistic radiation hydrodynamics beyond the effective description adopted here. In addition, uncertainties in $f_{\rm acc}$ propagate into the mapping between $M_i$ and $M_f$ and the interpretation of microlensing bounds.

Finally, even if alternative PBH formation mechanisms were to produce lighter objects that remain within the higher-dimensional regime, $M_{\rm PBH} < M_{\rm 4D}$, until matter-radiation equality, the framework developed here provides the tools necessary to analyze their phenomenology.
More broadly, this scenario opens a concrete observational pathway for testing extra-dimensional physics through macroscopic astrophysical signatures.

\section{Conclusions}
\label{sec:conclusions}

In this work, we have studied the coupled cosmological evolution of PBHs and radiation in the Arkani-Hamed-Dimopoulos-Dvali framework with large extra dimensions. By following the interplay between Hawking evaporation, radiation accretion, and cosmological expansion, we have shown that the modified black-hole geometry in the higher-dimensional regime can qualitatively alter the standard PBH evolution. In particular, the enlarged horizon radius and suppressed Hawking temperature can trigger a phase of runaway accretion, allowing initially microscopic PBHs to grow efficiently and potentially reach macroscopic, even solar-mass, scales by matter-radiation equality, for sufficiently large accretion efficiencies and early formation times. A central result of our analysis is that PBHs can account for the observed DM density with initial abundances as small as $\beta \sim 10^{-44}$, far below the values typically required in standard four-dimensional cosmology, where $\beta \sim 10^{-8}$--$10^{-20}$ is more common~\cite{Carr:2020gox, Villanueva-Domingo:2021spv}. In our scenario, this follows from the large mass amplification factor acquired during the early growth phase, so that the final PBH abundance is enhanced according to the scaling $\Omega_{\rm PBH} \sim \beta\, M_{\rm PBH}/M_i$~\cite{DeLuca:2020fpg}. Extra-dimensional effects open a previously unexplored growth-based pathway through which initially microscopic PBHs can become cosmologically relevant DM candidates.

The transition at the critical mass $M_{\rm 4D}$ plays a central role in this picture, since it marks the point at which PBHs cease to evolve as higher-dimensional objects and begin to follow standard four-dimensional Schwarzschild dynamics. This transition controls both the duration of the enhanced accretion phase and the final PBH mass spectrum, and is essential for determining the viable DM parameter space in ADD scenarios. At the same time, our results rely on an effective treatment of accretion in the early Universe. In particular, the efficiency of the growth phase depends on the accretion parameter $f_{\rm acc}$ and on the maximum temperature of the radiation bath, and is sensitive to the reheating history after inflation. Low reheating temperatures could reduce, or even eliminate, the period during which accretion dominates over evaporation. More generally, our treatment should be regarded as an effective estimate of the growth dynamics, rather than a first-principles calculation of relativistic accretion in a $(4+n)$-dimensional plasma background. A more complete analysis, including relativistic radiation hydrodynamics and extended PBH mass functions, would be needed to sharpen the quantitative predictions. 

Our results can be compared with recent investigations of thermal absorption and enhanced PBH mass growth in standard 4D scenarios, such as the frameworks discussed in Refs.~\cite{Das:2025vts,Kalita:2025fcs,Kallifatides:2026sik,Haque:2026vvp}. A fundamental difference lies in the conditions required to trigger the runaway accretion phase. In the 4D setup of Refs.~\cite{Haque:2026vvp,Kallifatides:2026sik}, unbounded mass growth occurs only above a critical collapse efficiency parameter, $\gamma \ge \gamma_c \simeq 0.33$. Conversely, in our ADD framework, the modified mass-radius relation allows a runaway accretion phase to occur even for the conventional value $\gamma = 0.2$ adopted here. For $n \ge 2$, the LED framework yields a significantly enlarged horizon radius $r_h$ and a suppressed Hawking temperature at a fixed mass compared to the 4D case. This geometric enhancement increases radiation absorption by many orders of magnitude, allowing initially microscopic PBHs with $M_i \gtrsim 10^{12}$\,g to grow past the solar mass threshold $M \gtrsim M_\odot$ before matter-radiation equality. Such a macroscopic amplification is significantly more challenging to achieve within standard cosmology. It would be interesting to further explore the dependence of this mechanism on the collapse efficiency parameter $\gamma$. A larger value of $\gamma$ shifts PBH formation to an earlier epoch and a hotter radiation bath, widening the period over which accretion is efficient and potentially enhancing the resulting PBH mass growth. Furthermore, Ref.~\cite{Das:2025vts} derives an accretion formula based on spherical accretion onto a Schwarzschild BH, incorporating specific relativistic corrections. A similar approach is adopted in Ref.~\cite{Kalita:2025fcs} for the Kerr solution, demonstrating that accretion is enhanced for low spin parameters.

Our framework might have relevant cosmological implications in light of recent observations by the James Webb Space Telescope, which have revealed an unexpectedly abundant population of bright galaxies and active galactic nuclei at high redshifts~\cite{Bogdan:2023ilu, Maiolino:2023zdu, Iocco:2024rez}. The existence of massive BHs and galaxies at very early times may challenge conventional formation scenarios, in which stellar mass seeds are required to undergo rapid accretion over a relatively short cosmic time; see, e.g., Ref.~\cite{Carr:2020erq}. In our ADD scenario, the enhanced horizon radius and suppressed Hawking temperature can trigger an early phase of efficient mass growth during the radiation-dominated era, amplifying initially microscopic PBHs by several orders of magnitude well before matter-radiation equality. Consequently, heavy PBH seeds can be generated at very high redshifts in this framework, potentially alleviating some of the tension associated with the formation of high-redshift SMBHs. A quantitative assessment of this possibility would require a dedicated analysis of the PBH seed mass function, their subsequent accretion history, and merger evolution, which we leave for future work.

Nevertheless, the results presented here demonstrate that extra dimensions can qualitatively reshape PBH cosmology and provide a novel mechanism for generating macroscopic DM from microscopic primordial seeds.

\begin{acknowledgments}
GFV, GL, and LV acknowledge support by Istituto Nazionale di Fisica Nucleare (INFN) through the Commissione Scientifica Nazionale 4 (CSN4) Iniziativa Specifica ``Quantum Universe'' (QGSKY). TKP is supported by Science and Technology Facilities Council (STFC) under ST/X003167/1. This article is based on the work from COST Actions COSMIC WISPers CA21106 and BridgeQG CA23130, supported by COST (European Cooperation in Science and Technology).
\end{acknowledgments}

\appendix

\section{Dependence on the Accretion Efficiency}
\label{app:facc_dependence}

In this appendix, we complement the discussion in Sec.~\ref{sec:discussion} by showing how the observational constraints on the primordial collapse fraction $\beta(M_i)$ behave under different choices of the accretion efficiency. As stated in the main text, the constraints in Fig.~\ref{fig:beta_cnstrnt} are computed for a benchmark value of $f_{\rm acc} = 10^{-3}$. We display the corresponding remapped constraints for $f_{\rm acc} = 10^{-7}$, see Fig.~\ref{fig:beta_cnstrnt1em7}, and $f_{\rm acc} = 10^{-1}$, see Fig.~\ref{fig:beta_cnstrnt1em1}. As $f_{\rm acc}$ decreases, the early runaway accretion phase becomes less efficient due to a reduced mass growth rate. Consequently, the remapping effect $M_i \rightarrow M_f$ is less pronounced, shifting the excluded regions back toward larger initial masses. This significantly expands the allowed parameter space where microscopic primordial black holes can safely account for the totality of the observed dark matter abundance without violating current microlensing bounds. Conversely, for a larger efficiency $f_{\rm acc} = 10^{-1}$, the runaway growth is accelerated, and the observational constraints compress the remaining window even further.

\begin{figure*}[htb!]
    \centering
    \subfigure{
        \includegraphics[height=7.5cm]{cnstrntn2facc10_-7.png}
    }\hfill
    \subfigure{
        \includegraphics[height=7.5cm,trim=1 0 0 0,clip]{cnstrntn2Mstar2facc10_-7.png}
    }\\[2mm]
    \subfigure{
        \includegraphics[height=7.5cm]{cnstrntn3facc10_-7.png}
    }\hfill
    \subfigure{
        \includegraphics[height=7.5cm,trim=1 0 0 0,clip]{cnstrntn4facc10_-7.png}
    }
    \caption{Same as Fig.~\ref{fig:beta_cnstrnt}, for $f_{\rm acc} = 10^{-7}$.}
    \label{fig:beta_cnstrnt1em7}
\end{figure*}

\begin{figure*}[htb!]
    \centering
    \subfigure{
        \includegraphics[height=7.5cm]{cnstrntn2facc10_-1.png}
    }\hfill
    \subfigure{
        \includegraphics[height=7.5cm,trim=1 0 0 0,clip]{cnstrntn2Mstar2facc10_-1.png}
    }\\[2mm]
    \subfigure{
        \includegraphics[height=7.5cm]{cnstrntn3facc10_-1.png}
    }\hfill
    \subfigure{
        \includegraphics[height=7.5cm,trim=1 0 0 0,clip]{cnstrntn4facc10_-1.png}
    }
    \caption{Same as Fig.~\ref{fig:beta_cnstrnt}, for $f_{\rm acc} = 10^{-1}$.}
    \label{fig:beta_cnstrnt1em1}
\end{figure*}

\clearpage
\bibliography{references.bib}
\end{document}